\pdfoutput=1
\documentclass[11pt]{article}

\usepackage{graphicx}
\usepackage{amsmath}
\usepackage{amssymb}
\usepackage{dcolumn}
\usepackage{bm}
\usepackage{slashed}
\usepackage{authblk}
\usepackage{cite}

\newcommand{\be}{\begin{equation}}
\newcommand{\ee}{\end{equation}}
\newcommand{\ba}{\begin{eqnarray}}
\newcommand{\ea}{\end{eqnarray}}
\newcommand{\no}{\nonumber \\}
\newcommand{\gsim}{\mathrel{\hbox{\rlap{\lower.55ex \hbox {$\sim$}}
                   \kern-.3em \raise.4ex \hbox{$>$}}}}
\newcommand{\lsim}{\mathrel{\hbox{\rlap{\lower.55ex \hbox {$\sim$}}
                   \kern-.3em \raise.4ex \hbox{$<$}}}}

\def\vk{{\vec k}}
\def\vq{{\vec q}}
\def\vp{{\vec p}}

\def\vj{{\vec j}}
\def\vE{{\vec E}}
\def\vB{{\vec B}}

\def\roughly#1{\mathrel{\raise.3ex\hbox{$#1$\kern-.75em%
\lower1ex\hbox{$\sim$}}}}
\def\lsim{\roughly<}
\def\gsim{\roughly>}

\def\hb{\hbar}
\def\({\left(}
\def\){\right)}
\def\[{\left[}
\def\]{\right]}

\def\l{{\lambda}}
\def\L{{\Lambda}}
\def\d{{\delta}}

\def\e{{\epsilon}}

\def\a{{\alpha}}
\def\b{{\beta}}
\def\c{{\chi}}

\def\g{{\gamma}}
\def\G{{\Gamma}}
\def\h{\eta}
\def\p{{\pi}}
\def\P{{\Pi}}
\def\m{{\mu}}
\def\n{{\nu}}
\def\r{{\rho}}
\def\s{{\sigma}}
\def\S{{\Sigma}}

\def\th{{\theta}}

\def\ps{{\psi}}

\def\x{{\xi}}
\def\P{{\Pi}}

\newcommand{\lag}{\langle}
\newcommand{\rag}{\rangle}

\newcommand{\pd}{{\partial}}
\newcommand{\dg}{\dagger}
\newcommand{\pr}{\parallel}
\newcommand{\ind}{\text{ind}}
\newcommand{\pb}{\text{PB}}
\newcommand{\tr}{\text{tr}}
\newcommand{\pp}{\perp}

\date{\today}

\begin{document}

\title{\bf Quantum Kinetic Theory for Quantum Electrodynamics}

\author[]{Shu Lin
\thanks{linshu8@mail.sysu.edu.cn}}
\affil[]{School of Physics and Astronomy, Sun Yat-Sen University, Zhuhai 519082, China}

\maketitle

\begin{abstract}

We derive a quantum kinetic theory for QED based on Kadanoff-Baym equations for Wigner functions. By assuming parity invariance and considering a complete set of self-energy diagrams, we find the resulting kinetic theory expanded to lowest order in $\hb$ generalizes the well-known classical kinetic theory to massive case. It contains elastic and inelastic collision terms and integrates screening effect naturally. For a given solution to the classical kinetic theory, we find at next order in $\hb$ a non-dynamical quantum correction to Wigner functions for both fermions and photons, which gives rise to spin polarization for fermion and photon respectively. The approach allows us to study the non-dynamical part of collisional effect on spin polarization phenomenon.

\end{abstract}


\newpage

\section{Introduction and Summary}

Spin polarization phenomena have been observed in a variety of experiments in particle physics \cite{STAR:2017ckg,STAR:2018gyt,STAR:2019erd} and condensed matter physics \cite{PhysRevLett.119.077202,PhysRevB.87.180402}. It can be induced by different sources such as electromagnetic field \cite{Metlitski:2005pr,Son:2004tq}, vorticity \cite{Erdmenger:2008rm,Banerjee:2008th,Neiman:2010zi,Landsteiner:2011cp}. In order to describe spin transports in a non-equilibrium setting, two complementary frameworks have been developed. One is spin kinetic theory \cite{Hattori:2019ahi,Weickgenannt:2019dks,Gao:2019znl,Liu:2020flb,Guo:2020zpa}, which generalizes the chiral kinetic theory \cite{Son:2012wh,Son:2012zy,Stephanov:2012ki,Pu:2010as,Chen:2012ca,Hidaka:2016yjf,Manuel:2013zaa,Manuel:2014dza,Huang:2018wdl,Carignano:2018gqt,Gao:2018wmr,Wang:2019moi,Lin:2019ytz,Gao:2019zhk,Hayata:2020sqz,Yang:2020mtz,Lin:2019fqo,Lin:2021sjw,Luo:2021uog} to massive case. The spin kinetic theory is organized as a systematic expansion in $\hb$, with the zeroth order gives the classical kinetic theory and spin effect appears from the first order. Recently much progress has been made towards realistic collisions term for the spin kinetic theory \cite{Hidaka:2016yjf,Zhang:2019xya,Li:2019qkf,Carignano:2019zsh,Yang:2020hri,Wang:2020pej,Shi:2020htn,Weickgenannt:2020aaf,Hou:2020mqp,Yamamoto:2020zrs,Weickgenannt:2021cuo,Sheng:2021kfc,Wang:2021qnt}. However a complete derivation of realistic collision term for QCD has yet to be performed.

The other framework is spin hydrodynamics \cite{Florkowski:2017ruc,Montenegro:2017rbu,Montenegro:2018bcf,Hattori:2019lfp,Gallegos:2021bzp,Florkowski:2018fap,Becattini:2018duy,Bhadury:2020puc,Fukushima:2020ucl,Li:2020eon,Hongo:2021ona,Peng:2021ago,She:2021lhe}, which includes spin density as an additional degree of freedom in conventional hydrodynamics. It is assumed that the spin density relaxes much slower than the other non-hydrodynamic modes. The assumption clearly depends on particular microscopic theory. It is desirable to confirm that such an assumption is indeed satisfied in systems of our interest.

In this paper, we derive quantum kinetic equations for quantum electrodynamics (QED) based on Kadanoff-Baym equations. This approach allows us to incorporate complete collision term for fermions and photons. By performing an expansion in $\hb$ of the resulting kinetic equations, we obtain at the lowest order a classical kinetic theory.
On general ground, one expects the classical kinetic theory to agree with the classical Boltzmann equations written down by Arnold, Moore and Yaffe (AMY) \cite{Arnold:2002zm,Arnold:2000dr,Arnold:2003zc} for quantum chromodynamics (QCD), which is essentially spin-averaged kinetic equations.
The agreement is achieved by assumption of parity invariance. It is known that for massive fermion, the kinetic theory contains the following degrees of freedom: $f^e_V/f^e_A$, $a_\m$ being vector/axial charge distribution functions and spin direction respectively \cite{Hattori:2019ahi}. By assuming the system is parity invariant, we are left with $f^e_V$ (to be denoted as $f_e$) as the only degree of freedom. A similar reduction applies to the photonic sector of kinetic theory, leaving only the distribution function of unpolarized photon $f_\g$ as the degree of freedom. With these simplifications, we find the resulting classical kinetic theory generalizes the classical Boltzmann equation by AMY \cite{Arnold:2002zm,Arnold:2000dr,Arnold:2003zc} to massive case for QED. The generalization to massive case is crucial for the phenomenology of spin polarization.

The classical kinetic equation serves as a background, on which we can study quantum correction by systematic expansion in $\hb$ within our approach. As we show in Section \ref{sec_hbar}, the quantum correction at next order gives rise to spin polarization of both fermions and photons. Our approach bridges the gap between realistic classic kinetic theory and formal development of quantum kinetic theory, thus allowing us to study collisional contribution to spin polarization phenomena in a realistic system. It is also worth mentioning that in early studies of transport coefficients in classical kinetic theory \cite{Arnold:2002zm,Arnold:2000dr,Arnold:2003zc}, gradient expansion is employed while $\hb$ is set to unity. In fact, we shall clarify the relation between two expansions in Section \ref{sec_hbar}. From our approach, it is manifest that gradient expansion in classical kinetic theory already includes partial contribution of $\hb$ expansion, which does not lead to spin polarization. We shall still use the terminology classical and quantum kinetic theory as a separation between spin unpolarized and polarized sectors of the kinetic theory.

It is informative to compare the classical kinetic theories for QED and QCD. On one hand, they share similar scales: characteristic momentum of quasi-particles $\L$; thermal/screening mass $e\L(g\L)$; damping rate of quasi-particles $e^2\L(g^2\L)$. One the other hand, they differ in one fundamental aspect: QED does not possess a non-pertubative scale while QCD does. The non-perturbative scale for QCD cuts off long range fluctuation of chromoelectromagnetic fields, allowing for omission fluctuation of chromoelectromagnetic fields beyond the scale $\frac{1}{g^2\L}$. However such a mechanism is not present in QED. We will instead choose a larger coarse-graining scale $\frac{1}{e^4\L}$ for the kinetic theory and restrict ourselves to the situation without background electromagnetic fields. We will show electromagnetic fields generated by fluctuations at the scale $\frac{1}{e^4\L}$ have a subleading effect as compared to that of spacetime gradients.


The paper is organized as follows: in Section \ref{sec_KB}, we derive the Kadanoff-Baym equations for fermions and photons. By the assumption of parity invariance, the degrees of freedom are identified as unpolarized fermion/photon distribution functions. We also discuss regimes of validity of the kinetic theory; in Sections \ref{sec_ec} and \ref{sec_inec}, we elaborate on the self-energies of fermions and photons, which gives rise to the elastic and inelastic collision terms. In order to incorporate the interference term in $2\leftrightarrow2$ process and $1\leftrightarrow2$ process, it is crucial to include vertex correction in the self-energies. The resulting kinetic equations to lowest order in $\hb$ shows a clear resemblance to the Boltzmann equations by AMY; in Section \ref{sec_hbar}, we consider the next order expansion in $\hb$. After enumerating possible $\hb$ corrections, we will focus on a non-dynamical type of correction, which gives rise to spin polarization and is entirely fixed by solution of classical kinetic equation; we provide outlook in Section \ref{sec_outlook}.

\section{Kadanoff-Baym Equations for QED}\label{sec_KB}

\subsection{Fermionic Kadanoff-Baym Equations}

We begin by writing down the fermionic part of the Lagrangian 
\begin{align}\label{Lag_f}
\frac{{\cal L}}{\hb}=\bar{\psi}\(i\slashed{\pd}-e\slashed{A}-\frac{m}{\hb}\)\psi+\bar{\h}\ps+\bar{\ps}\h,
\end{align}
where $\bar{\h}$ and $\h$ are sources coupled to $\ps$ and $\bar{\ps}$ respectively. We first write down the Dyson-Schwinger equation on the Schwinger-Keldysh (SK) contour
\begin{align}\label{DS_f}
&\(i\slashed{\pd}_x-\frac{m}{\hb}\)\ps(x)=\h(x)+\h_{\ind}(x),\\
&\bar{\ps}(y)\(-i\overleftarrow{\slashed{\pd}}_y-\frac{m}{\hb}\)=\bar{\h}(y)+\bar{\h}_{\ind}(y),
\end{align}
with $\h_\ind=e\slashed{A}\ps$ and $\bar{\h}_\ind=e\bar{\ps}\slashed{A}$\footnote{Recall the dimensions of fields: $\ps\sim\text{length}^{-3/2}$, $A\sim\text{length}^{-1}$. By writing down the vertex in \eqref{Lag_f}, we have redefined the coupling constant $e$ such that it contains no factor of $\hb$.}.
For simplicity, we have assumed the absence of background gauge field. 
Following \cite{Blaizot:2001nr}, we take the derivative $i\frac{\d}{\d\h(y)}$ on \eqref{DS_f} to obtain
\begin{align}\label{hh}
\(i\slashed{\pd}_x-\frac{m}{\hb}\)S_c(x,y)=i\d_c(x-y)+i\frac{\h_{\ind}(x)}{\h(y)},
\end{align}
where the subscript indicates the quantity being contour-time ordered. The matrix form of $S_c(x,y)$ in $12$ basis is given by
\begin{align}\label{Sc}
S_{c,\a\b}(x,y)
=\begin{pmatrix}
S_{\a\b}^{11}(x,y)& S_{\a\b}^<(x,y)\\
S_{\a\b}^>(x,y)& S_{\a\b}^{22}(x,y)
\end{pmatrix}
=\begin{pmatrix}
\lag T\ps_\a(x)\bar{\ps}_\b(y)\rag& -\lag\bar{\ps}_\b(y)\ps_\a(x)\rag\\
\lag\ps_\a(x)\bar{\ps}_\b(y)\rag& \lag\bar{T}\ps_\a(x)\bar{\ps}_\b(y)\rag
\end{pmatrix}.
\end{align}
We have also kept explicit Dirac indices $\a\b$ in \eqref{Sc}. $T$ and $\bar{T}$ correspond to time ordering and anti-time ordering. The last term of \eqref{hh} can be evaluated as
\begin{align}
\frac{\d}{\d\h(y)}\h_\ind(x)=\int d^4z\(-i\frac{\d}{\d\ps(z)}\h_\ind(x)\)\(i\frac{\d}{\d\h(y)}\d\ps(z)\)=\int d^4z\S_c(x,z)S_c(z,y),
\end{align}
with $\S_c(x,y)=-i\frac{\d}{\d\ps(y)}\h_\ind(x)=\lag\h_\ind(x)\bar{\h}_\ind(y)\rag_c$.
Taking $x$ and $y$ on the forward and backward contours respectively, we obtain
\begin{align}\label{KB_raw}
\(i\slashed{\pd}_x-\frac{m}{\hb}\)S^<(x,y)=i\int d^4z\S_F(x,z)S^<(z,y)-i\int d^4z\S^<(x,z)S_{\bar{F}}(z,y).
\end{align}
We can eliminate the time ordered self-energy $\S_F$ and anti-time ordered correlator $S_{\bar{F}}$ in favor of retarded/advanced correlator defined as
\begin{align}
S_R(x,y)&=i\th(x_0-y_0)\(S^>(x,y)-S^<(x,y)\)\no
S_A(x,y)&=-i\th(y_0-x_0)\(S^>(x,y)-S^<(x,y)\),\nonumber
\end{align}
and similar definitions for $\S_{R/A}$.
Using the following relations
\begin{align}\label{corr_rel}
&\S_F(x,z)=-i\S_R(x,z)+\S^<(x,z)\no
&S_{\bar{F}}(z,y)=S^<(z,y)+iS_A(z,y),
\end{align}
we arrive at the standard form of KB equations for fermions
\begin{align}\label{KB_f}
\(i\slashed{\pd}_x-\frac{m}{\hb}\)S^<(x,y)=\int d^4z\(\S_R(x,z)S^<(z,y)+\S^<(x,z)S_A(z,y)\).
\end{align}
We will rewrite \eqref{KB_f} in terms of Wigner transformed lesser propagator
\begin{align}
\tilde{S}^<(X=\frac{x+y}{2},P)=\int d^4(x-y)e^{iP\cdot (x-y)/\hb}\lag S^<(x,y)\rag.
\end{align}
The RHS can be expressed using $\tilde{S}^<(X,P)$ and counterpart for self-energy by the following expansion
\begin{align}\label{grad_exp}
\int d^4zA(x,z)B(z,y)=\int_K e^{-iK\cdot(x-y)}\(\tilde{A}\tilde{B}+\frac{i\hb}{2}\{\tilde{A},\tilde{B}\}_\pb\)+O(\hb^2),
\end{align}
where the Poisson bracket is defined as
\begin{align}\label{PB}
\{\tilde{A},\tilde{B}\}=\pd_k\tilde{A}\cdot\pd_X\tilde{B}-\pd_X\tilde{A}\cdot\pd_k\tilde{B},
\end{align}
and $\int_K\equiv\int\frac{d^4K}{(2\p)^4}$.
The Wigner transform of \eqref{KB_f} then reads
\begin{align}\label{KB_f_W}
\frac{i}{2}\slashed{\pd}S^<+\frac{\slashed{P}-m}{\hb}S^<=\(\S_RS^<+\S^<S_A\)+\frac{i\hb}{2}\(\{\S_R,S^<\}_\pb+\{\S^<,S_A\}_\pb\).
\end{align}
We have dropped the tildes and the common arguments $(X,P)$ for propagators and self-energies for notational simplicity.

To proceed further, we use the quasi-particle approximation \cite{Blaizot:2001nr}, in which the spectral density is given by
\begin{align}\label{qpa}
\r(X,P)=S^>(X,P)-S^<(X,P)=2\p\hb\e(p_0)\(\slashed{P}+m\)\d(P^2-m^2-2P^\m Re\S^R_\m),
\end{align}
where $\e(p_0)$ is the sign function and $\S_\m^R=\frac{1}{4}tr[\g_\m\S^R]$. The structure of \eqref{qpa} indicates quasi-particle has a momentum shifted by $Re\S^R_\m$ and has a vanishing damping rate $\G(X,P)$. For system with characteristic momenta $\L$, $Re\S^R\sim e\L$ and damping rate $\sim e^2\L$. We can indeed ignore $Re\S^R_\m$ next to $P_\m$ and $\G$.
We can use the following representations to further simplify \eqref{KB_f_W}\footnote{The real part is formally defined with hermitian conjugate as $ReA=\frac{1}{2}\(A+\g^0A^\dg\g^0\)$.}
\begin{align}
&\S^R=Re\S^R+\frac{i}{2}\(\S^>-\S^<\)\no
&S^A=ReS^R-\frac{i}{2}\(S^>-S^<\),\nonumber
\end{align}
and ignore $Re\S^R$ and $ReS^R\propto\G$ as reasoned above to arrive at
\begin{align}\label{KB_f_final}
\frac{i}{2}\slashed{\pd}S^<+\frac{\slashed{P}-m}{\hb}S^<=\frac{i}{2}\(\S^>S^<-\S^<S^>\)-\frac{\hb}{4}\(\{\S^>,S^<\}_\pb-\{\S^<,S^>\}_\pb\).
\end{align}
\eqref{KB_f_final} can be solved order by order in $\hb$
\begin{align}\label{S_hbar}
S^<=S^{<(0)}+\hb S^{<(1)}+\hb^2 S^{<(2)}+\cdots,
\end{align}
with the lowest order solution given by \cite{Hattori:2019ahi}\footnote{The overall factor of $\hb$ can be removed by a redefinition of the Wigner function. We will give a more detailed account of the formal expansion in Sec.~\ref{sec_hbar}.}
\begin{align}\label{S0_zm}
S^{<(0)}=-2\p\hb \e(P\cdot u)\d(P^2-m^2)\((\slashed{P}+m)f^e_V+\g^5\slashed{a}f^e_A-\frac{\s^{\m\n}}{2}\frac{\e_{\m\n\r\s}P^\r a^\s}{m}f^e_A\).
\end{align}
Note that we have introduced an observer's frame vector $u$ in the sign function, which separates particles and anti-particles. Despite appearance of $u$, it is actually frame independent at this order because boosts do not change a particle into anti-particle. $f^e_V/f^e_A$ and $a^\m$ correspond to vector/axial distributions and spin direction vector respectively. We will assume that the system is parity invariant, which means $f^e_A=0$, leaving $f^e_V$ as the only degree of freedom.
\footnote{It is parity invariant in the sense that there is no difference between distributions of opposite chiral components and similarly for photons. However, in an out-of-equilibrium setting sources with both parities such as spacetime gradient of temperature and fluid velocity etc.}
This is identified as fermion distribution function in classical kinetic theory. We will denote $f^e_V$ as $f_e$ below. The lowest order solution can be written as
\begin{align}\label{S0}
S^{<(0)}(X,P)=-2\p\hb \e(P\cdot u)\d(P^2-m^2)(\slashed{P}+m)f_e(X,P),
\end{align}
which is nothing but the equilibrium fermion propagator with Fermi-Dirac distribution promoted to spacetime dependent distribution $f_e$.

In fact $f_e$ satisfies the constraint $f_e(X,P)+f_e(X,-P)=1$ like Fermi-Dirac distribution. To see the origin of the constraint, we can show from the definition \eqref{Sc} that $\tr[S^<(x,y)]=-\tr[S^>(y,x)]$, which gives $\tr[S^<(X,P)]=-\tr[S^>(X,-P)]$. Combining with \eqref{S0}, we easily arrive at the constraint. Furthermore, we have the following property
\begin{align}
S^{>(0)}(X,-P,m)=S^{<(0)}(X,P,-m).
\end{align}
Defining
\begin{align}\label{S0overline}
\overline{S}^{>/<(0)}(X,P,m)={S}^{>/<(0)}(X,P,-m),
\end{align}
we can then write down the following useful relations
\begin{align}\label{reciprocal_f}
\overline{S}^{<(0)}(X,P,m)=S^{>(0)}(X,-P,m),
\end{align}
and a similar one with $>\leftrightarrow<$. Below $S$ and $\overline{S}$ always contain the argument $m$, which we omit from now. \eqref{reciprocal_f} suggests the interpretation of $\overline{S}^{>/<(0)}$ as charge-conjugated Wigner function: To arrive at the classical kinetic theory, we will always deal with fermions with positive energy. For $P\cdot u>0$, we use $S^{<(0)}(X,P)$ to describes particles, while for $P\cdot u<0$, we use $\overline{S}^{>(0)}(X,-P)$ instead to switch to anti-particle description.

\subsection{Photonic Kadanoff-Baym Equations}

We now turn to the photonic counterpart starting with the following relevant part of the Lagrangian
\begin{align}\label{Lag_p}
\frac{{\cal L}}{\hb}=-\frac{1}{4}F_{\m\n}^2+\bar{\ps}\(-e\slashed{A}_\m\)\ps-A_\m j^\m-\frac{1}{2\x}\(P^{\a\b}\pd_\a A_\b\)^2,
\end{align}
where $j^\m$ is the source to $A_\m$. The last term is the gauge fixing term for Coulomb gauge, with the projector $P^{\a\b}$ defined with an observer's frame vector $n^\a$ as: $P^{\a\b}=n^\a n^\b-g^{\a\b}$. We will take $n^\m$ to be the same frame vector as the fermionic one $u^\m$. It projects out the temporal component of a vector. The Coulomb gauge singles out transversely polarized photon as the physical degrees of freedom, offering a quick path to kinetic theory.

The Dyson-Schwinger equation for photon is given by
\begin{align}\label{photon_DS}
\(\pd^2g^{\m\n}-\pd^\m\pd^\n\)A_\n(x)+\frac{1}{\x}\(P^{\m\n}\pd_\n P^{\a\b}\pd_\a A_\b(x)\)=j^\m(x)+j_\ind^\m(x),
\end{align}
where $j_\ind^\m=\bar{\ps}\g^\m\ps$. Taking the derivative $i\frac{\d}{\d j^\r(y)}$ on the SK-contour, we obtain
\begin{align}\label{KB_p}
\(\pd^2g^{\m\n}-\pd^\m\pd^\n+\frac{1}{\x}P^{\m\a}P^{\n\b}\pd_\a\pd_\b\)_xD^c_{\n\r}(x,y)=i\d^\m_\r\d_c(x-y)+i\int d^4z\P_c^{\m\n}(x,z)D_{\n\r}^c(z,y),
\end{align}
where $D^c_{\n\r}(x,y)=\lag A_\n(x)A_\r(y)\rag_c$ and $\P_c^{\m\n}(x,y)=\lag j_\ind^\m(x)j_\ind^\n(y)\rag_c$. The subscript $x$ indicates the derivatives acting on $x$. The usual Coulomb gauge is recovered in the limit $\x\to0$:
\begin{align}
P^{\m\a}\pd_\a^x D_{\m\n}^<(x,y)=0.
\end{align}
Following similar procedures as the fermionic case for the Wigner transform and quasi-particle approximation, we arrive at
\begin{align}\label{KB_p_final}
&\bigg[\frac{1}{\hb^2}\(-P^2g^{\m\n}+P^\m P^\n-\frac{1}{\x}P^{\m\a}P^{\n\b}P_\a P_\b\)+\frac{i}{2\hb}\bigg(-2P\cdot\pd g^{\m\n}+(\pd^\m P^\n+\pd^\n P^\m) \no
&-\frac{1}{\x}P^{\m\a}P^{\n\b}(\pd_\a P_\b+\pd_\b P_\a)\bigg)+\frac{1}{4}\(\pd^2g^{\m\n}-\pd^\m\pd^\n+\frac{1}{4\x}P^{\m\a}P^{\n\b}\pd_a\pd_\b\)\bigg]D^<_{\n\r}=\no
&\frac{i}{2}\(\P^{\m\n>}D_{\n\r}^<-\P^{\m\n<}D_{\n\r}^>\)+\frac{\hb}{4}\{\P^{\m\n>},D_{\n\r}^<\}-\frac{\hb}{4}\{\P^{\m\n<},D_{\n\r}^>\},
\end{align}
with the Coulomb gauge condition
\begin{align}\label{gauge}
P^{\m\a}\(\frac{\hb}{2}\pd_\a-iP_\a\)D^<_{\m\n}=0.
\end{align}
We seek solution to \eqref{KB_p_final} and \eqref{gauge} order by order in $\hb$
\begin{align}\label{D_hb}
D_{\m\n}^<=D_{\m\n}^{<(0)}+\hb D_{\m\n}^{<(1)}+\cdots,
\end{align}
with the lowest order solution given by \cite{Hattori:2020gqh,Huang:2020kik,Huang:2018aly}
\begin{align}\label{D0_zm}
D_{\m\n}^{<(0)}=2\p\hb^2\e(P\cdot u)\d(P^2)\(P_{\m\n}^Tf^\g_V-iS_{\m\n}f^\g_A\),
\end{align}
where $P_{\m\n}^T=P_{\m\n}-\frac{P_{\m\a}P_{\n\b}P^\a P^\b}{-P^2+(P\cdot u)^2}$ and $S_{\m\n}=\frac{\e_{\m\n\r\s}P^\r u^\s}{P\cdot u}$ are parity even and odd projectors perpendicular to $P_{\m\a}P^\a$. With these projectors, the Coulomb gauge condition \eqref{gauge} is automatically satisfied. Similar to the fermionic case, in a parity invariant system, $f^\g_A=0$ and we denote $f^\g_V$ as $f_\g$, which is to be identified as photon distribution function. We have then a simplified Wigner function for photon
\begin{align}\label{D0}
D_{\m\n}^{<(0)}(X,P)=2\p\hb^2\e(P\cdot u)\d(P^2)P_{\m\n}^T f_\g(X,P),
\end{align}
which is nothing but the equilibrium photon propagator in Coulomb gauge with Bose-Einstein distribution promoted to spacetime dependent distribution $f_\g$. Similar to the fermionic case, the distribution satisfies the constraint $f_\g(X,-P)=-1-f_\g(X,P)$. It follows from
\begin{align}\label{reciprocal_p}
D^{<}_{\m\n}(x,y)=D^{>}_{\n\m}(y,x)\Rightarrow D_{\m\n}^{<(0)}(X,P)=D_{\n\m}^{>(0)}(X,-P)=D_{\m\n}^{>(0)}(X,-P),
\end{align}
upon using \eqref{D0}.

\subsection{Classical kinetic equations and regime of validity}

Now we determine the dynamics of $f_e$ and $f_\g$. For the former, it is known that $S^{<(1)}$ contains only axial and tensor components \cite{Yang:2020hri,Hattori:2019ahi}. Taking the trace of \eqref{KB_f_final}, we find $\tr[(\slashed{P}-m)S^{<(1)}]$=0, and we arrive at
\begin{align}\label{reduced_f}
&\tr[\slashed{\pd}S^{<(0)}]=\tr[\S^{>(0)}S^{<(0)}-\S^{<(0)}S^{>(0)}],
\end{align}
where $S^{<(0)}$ is given by \eqref{S0}.

For the latter, we assume $D_{\n\r}^{<(1)}$ is on-shell. The dynamics of $f_\g$ can be derived by contracting \eqref{KB_p_final} with $P_{\m}^{T,\r}$. We find $P_\m^{T,\r}\big(-P^2g^{\m\n}+P^\m P^\n-\frac{1}{\x}P^{\m\a}P^{\n\b}P_\a P_\b\big)D_{\n\r}^{<(1)}=0$ at $O(\hb^{-1})$. The remaining terms give the dynamical equation for $f_\g$:
\begin{align}\label{reduced_p}
&-2P\cdot\pd g^{\m\n}D_{\n\m}^{<(0)}=\hb\(\P^{\m\n>(0)}D_{\n\m}^{<(0)}-\P^{\m\n<(0)}D_{\n\m}^{>(0)}\).
\end{align}
The presence of $\hb$ in the classical kinetic equation is consistent with the dimension of photon self-energies $\P^{\m\n>/<}(X,P)\sim \text{length}^{-2}$.

Let us express the LHS of \eqref{reduced_f} and \eqref{reduced_p} in terms of distributions $f_e$ and $f_\g$. Taking $p_0>0$, we have
\begin{align}\label{KE_LHS}
&\tr[\slashed{\pd}S^{<(0)}(X,P)]=(-2\p\hb)4P\cdot\pd \d(P^2-m^2)f_{e}(X,P)=(-2\p\hb)\frac{2}{E_p}P\cdot\pd f_e(X,P) \d(p_0-E_p),\no
&-2P\cdot\pd g^{\m\n}D_{\n\m}^{<(0)}(X,P)=(2\p\hb^2)4P\cdot\pd \d(P^2)f_{\g}(X,P)=(2\p\hb^2)\frac{2}{p}P\cdot\pd f_\g(X,P) \d(p_0-p),
\end{align}
with $E_p=(p^2+m^2)^{1/2}$. 
Dividing out the xfactor $2$ converts the spin-summed collision term to the spin-averaged one for either fermion or photon. We have only kept the particle contributions in \eqref{KE_LHS}. The kinetic equation for anti-particles can be obtained from the corresponding equations for $S^{<(0)}(X,-P)$ and $D_{\n\r}^{<(0)}(X,-P)$.
Note that photon is its own anti-particle, so the resulting equation is expected to be equivalent. For fermions, we obtain
\begin{align}\label{KE_anti}
&\tr[\slashed{\pd}S^{<(0)}(X,-P)]=(2\p\hb)4P\cdot\pd \d(P^2-m^2)f_{e}(X,-P)=(-2\p\hb)\frac{2}{E_p}P\cdot\pd f_{\bar{e}}(X,P) \d(p_0-E_p),
\end{align}
with $f_{\bar{e}}(X,P)\equiv 1-f_e(X,-P)$ identified as distribution function for anti-particles. Note that $f_e(X,P)$, $f_{\bar{e}}(X,P)$ and $f_\g(X,P)$ all have positive energies, so we may also denote them as $f_e(X,\vp)$, $f_{\bar{e}}(X,\vp)$ and $f_\g(X,\vp)$.

In order to close the equations, we need to express $\S^{</>(0)}$ and $\P^{\m\n>/<(0)}$ on the RHS of \eqref{reduced_f} and \eqref{reduced_p} in terms of $f_e$ and $f_\g$ as well. This is the subject of the next two sections. As we shall see, the self-energies consistently incorporate elastic and inelastic collisions in the known classical kinetic theory. The final expressions for the elastic and inelastic contributions can be found in \eqref{loss_f}, \eqref{loss_p} and \eqref{coll_f}, \eqref{coll_p} respectively. We will set $\hb=1$ in the next two sections, with the understanding that factors of $\hb$ can always be reinstated by dimension in the classical collision term. We will retain $\hb$ in Section~\ref{sec_hbar} when we discuss quantum corrections.

Before presenting details on the self-energies, we discuss the regime of validity of the kinetic theory. The key conditions to be satisfied are the following:

i. A separation of scales between quasi-particle momenta $\L$, thermal mass $e\L$ and damping rate $e^2\L$;

ii. The physical observable of our interest is dominated by the dynamics of quasi-particles, which are described by the kinetic theory;

iii. The coordinate in the kinetic theory is coarse-grained within the scale $1/e^4\L$. Collisions are local on every coarse-grained spacetime point. The coarse-graining scale will also be crucial for condition v below;

iv. The distributions can only be weakly anisotropic such that instability associated with electromagnetic fields does not affect the dynamics of quasi-particles;

v. A background electromagnetic field is excluded by assumption. The effect of electromagnetic fields from thermal fluctuations is also neglected compared to spacetime derivatives of distributions $e|A|\ll \pd_Xf/f$. This is possible if we assume $e^6\L\ll \pd_Xf/f\lesssim e^4\L$. As we show below, the large conductivity of the medium suppresses the fluctuation of electromagnetic fields $e|A|\sim e^6\L$ so that our assumption on the gradient of distribution always dominates over electromagnetic fields from fluctuations;

vi. Since the equations and solutions are organized by expansion in $\hb$, we need to have the quantum correction small compared to the classical counterpart. This is guaranteed by $\hb\pd_Xf\ll \L$.

We will not elaborate on iv through vi below:

iv. It is known that anisotropic distribution in momentum can lead to filamentation instability, in which electromagnetic field draws energy from qausi-particles and grow exponentially \cite{Mrowczynski:1996vh}, see \cite{Mrowczynski:2016etf} for a recent review. The scale of the instability can be estimated following \cite{Romatschke:2003ms}. For a given class of anisotropic distributions $f_{\text{iso}}(\sqrt{{\bf p}^2+\x({\bf p}\cdot{\bf n})^2})$, with $\x$ and $n$ corresponding to magnitude and axis of ansiotropy respectively, the unstable modes for weakly anisotropic case $\x\ll1$ is found to have a characteristic freqeuency $\x^{3/2}e^2T$ \cite{Romatschke:2003ms}. In order for the unstable mode not to invalidate the kinetic theory, we require the instability occurs more slowly than characteristic time scale of the kinetic theory $\x^{3/2}e^2T\ll e^4T$, which leads to $\x^{3/2}\ll e^2$.

v. Now we justify the omission of electromagnetic fields. While we exclude background electromagnetic fields by assumption, the latter can still be generated by thermal fluctuations. We show that their effect is subleading compared to that of spacetime derivatives. The magnitude of the fluctuation depends on the property of the medium and the time scale of the fluctuation. For simplicity, we will probe the medium in equilibrium, which is expected to give the correct parametric estimate. We will probe the medium with an external current $(\r_{\text{ex}},\vj_{\text{ex}})$. The electromagnetic fields satisfy Maxwell equations
\begin{align}
&\nabla\times\vE=-\pd_t\vB,\quad \nabla\cdot\vB=0,\no
&\nabla\times{\vec H}=\vj_{\text{ex}}+\pd_t{\vec D},\quad \nabla\cdot{\vec D}=\r_{\text{ex}}.
\end{align}
The difference between ${\vec H}$ and $\vB$ is supposed to come from spin of charge carriers, which comes from next order in $\hb$, so we ignore. The electric displacement and electric field are related by ${\vec D}=\e \vE$, with $\e$ being the dielectric constant characterizing the property of the medium. We may solve ${\vE}$ in momentum space as:
\begin{align}\label{maxwell_q}
i\frac{\vq\times(\vq\times\vE)}{q_0}=\vj_{\text{ex}}-iq_0\e\vE.
\end{align}
\eqref{maxwell_q} encodes response of $\vE$ to $\vj_{\text{ex}}$. In the gauge $A_0=0$, \eqref{maxwell_q} gives rise to the following retarded correlator
\begin{align}\label{EA}
&\lag A_i(q_0,\vq)A_i(q_0',\vq')\rag_R\sim\frac{1}{q_0^2\e-q^2}\d(q_0+q_0')\d^3(\vq+\vq')\quad \vq\perp{\vec A},\no
&\lag A_i(q_0,\vq)A_i(q_0',\vq')\rag_R\sim\frac{1}{q_0^2\e}\d(q_0+q_0')\d^3(\vq+\vq')\quad \vq\parallel{\vec A}.
\end{align}
On the coarse-graining scale of the coordinate, the medium is known to be a very good conductor, so that we can parametrize the dielectric constant as
\begin{align}
\e\simeq1+\frac{i\s}{q_0},
\end{align}
with $\s\sim\frac{T}{e^2}$ \cite{Arnold:2000dr,Arnold:2003zc}. For the purpose of parametric estimation, we will regard $q\sim q_0$ and not distinguish between longitudinal and transverse cases in \eqref{EA}.
The retarded correlator in \eqref{EA} dictates the symmetrized correlator of electromagnetic fields by fluctuation-dissipation theorem \cite{Arnold:1996dy}
\begin{align}
\lag A_i(q_0,\vq)A_i(q_0',\vq')\rag_{sym}=\frac{2T}{q_0}Im\lag A_i(q_0,\vq)A_i(q_0',\vq')\rag_{R}\sim \frac{T}{q_0}Im\frac{1}{q_0^2+i\s q_0}\d(q_0+q_0')\d^3(\vq+\vq').
\end{align}
Noting that $\s\gg q_0$, we obtain the following estimate for fluctuation of electromagnetic gauge fields in coordinate space
%
\begin{align}\label{e4TA}
|A_i(t,x)|\sim\(\frac{T}{q_0^2\s}q_0 q^3\)^{1/2}\sim e^5T,
\end{align}
where we have Fourier transformed to coordinate space and taken $q_0\sim q\sim e^4T$ since we are concerned with fluctuation of electromagnetic field over the coarse-graining distance. It follows that the effect of fluctuating electromagnetic fields can be ignored because the spacetime derivatives are much larger than the mean fluctuation of electromagnetic fields $e^4T\gsim \pd_Xf/f\gg e|A_i|\sim e^6T$ by our assumption. The origin of the small fluctuation can be attributed to large conductivity, which disfavors fluctuation of electromagnetic fields.

While the electromagnetic field can be neglected over the coarse-graining scale, it does play a role within the scale $1/e\L$ in the form of exchanged virtual photon in fermion scatterings. The corresponding resummed propagator for virtual photon is fully determined by the distribution of real fermions in the kinetic theory, see details at the end of Sec.~\ref{sec_ec}. Similarly, the virtual fermion relevant for scattering between real fermion and photon is also fully determined by on-shell degree of freedoms in the kinetic theory.

vi. for the $\hb$ expansion to be valid, we require the quantum correction to be small. We shall obtain in Sec.~\ref{sec_hbar} that $\hb S^{<(1)}\sim \hb^2\pd_X f$, which is to be compared with $S^{<(0)}\sim \hb\L f$. This condition is clearly satisfied by $\hb\pd_Xf\ll \L$, i.e. the spacetime gradient needs to be small enough.

\section{Elastic collisions}\label{sec_ec}

We begin with the $2\to2$ elastic collisions. These can be obtained from two-loop contributions to self-energies. As we shall see, there are both corrections to propagators and vertices. The former give squares of amplitude in separate channels, and the latter give interference terms between different channels.

In this section and the next one, we work solely with zeroth order quantities: $S^{</>(0)}$, $\S^{</>(0)}$, $D_{\n\r}^{</>(0)}$, $\P^{\m\n</>(0)}$. To ease notations, we suppress the superscript $(0)$.

\subsection{propagator corrections}

\begin{figure}
\includegraphics[width=0.45\textwidth]{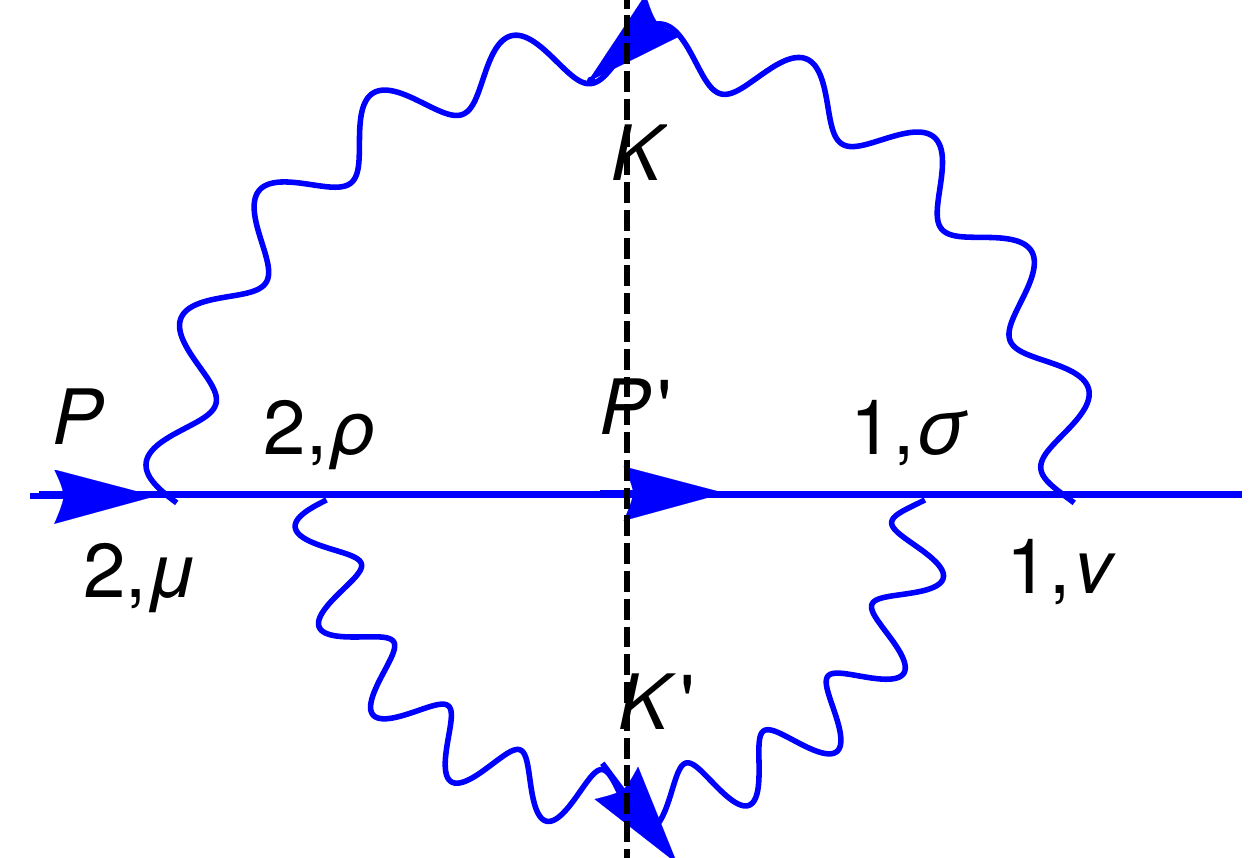}
\includegraphics[width=0.45\textwidth]{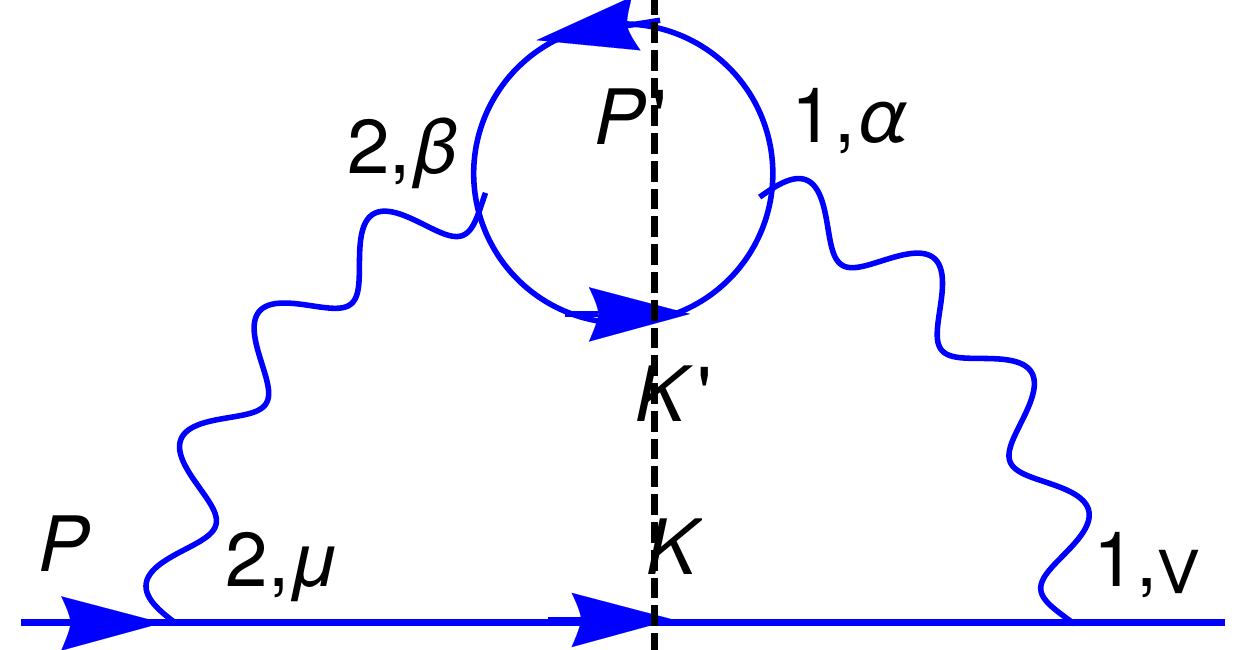}
\caption{\label{se-fermion}Two-loop diagrams for fermion self-energy containing propagator corrections. the $12$ labeling are uniquely determined by the requirement that three propagators attached to a vertex cannot be simultaneously on-shell. The on-shell particles are indicated by a cut (dashed line) through the corresponding propagators.}
\end{figure}

To be specific, we focus on the term $\S^>(P)S^<(P)$. $\S^>$ has the following representation
\begin{align}\label{Sigam}
\S^>(P)=e^2\int_K\g^\m S^>(K+P)\g^\n D_{\n\m}^<(K).
\end{align}
Since $P$ is on-shell, we cannot have both fermion momentum and photon momentum in the loop on-shell. Indeed, the one-loop contribution does not capture dissipation effects, for which two-loop diagrams are needed. The two-loop diagrams containing propagator corrections are shown in Fig.~\ref{se-fermion}, in which the left/right diagrams has one fermion/photon momentum off-shell. Note that the $12$ labels are uniquely determined by the requirement that three propagators attached to a vertex cannot be simultaneously on-shell.

The left panel of Fig.~\ref{se-fermion} gives
\begin{align}\label{SS}
\S^>(P)S^<(P)=e^4\int_{K,K'}\g^\m S_{22}(K+P)\g^\r S^>(P')\g^\s S_{11}(K+P)\g^\n D_{\n\m}^<(K)D_{\r\s}^>(K')S^<(P).
\end{align}
$\S^<(P)S^>(P)$ can be obtained from \eqref{SS} by the replacement $>\leftrightarrow <$ and $1\leftrightarrow 2$. The dependences of $S^{>/<}(P)$ on $f_e$ suggest that $\S^>(P)S^<(P)$ and $\S^<(P)S^>(P)$ correspond to loss and gain terms respectively. It is sufficient to focus on one of term only. 
We show in Appendix A that \eqref{SS} give rise to squares of s-channel Compton scattering, u-channel Compton scattering and t-channel annihilation.

The right panel of Fig.~\ref{se-fermion} gives
\begin{align}\label{SS2}
&\S^>(P)S^<(P)=-e^4\int_{P',K'}\g^\m S^>(K)\g^\n D_{\m\b}^{22}(P-K)D_{\a\n}^{11}(P-K)\tr[\g^\a S^<(P')\g^\b S^>(K')]\no
&\times S^<(P).
\end{align}
Again the replacement $>\leftrightarrow <$ and $1\leftrightarrow 2$ in the above leads to $\S^<(P)S^>(P)$.
We show in Appendix A that \eqref{SS2} give rise to squares of t-channel Coulomb scattering (between fermions), s-channel Coulomb scattering (between fermion and anti-fermion) and t-channel Coulomb scattering (between fermion and anti-fermion).

\begin{figure}
\includegraphics[width=0.45\textwidth]{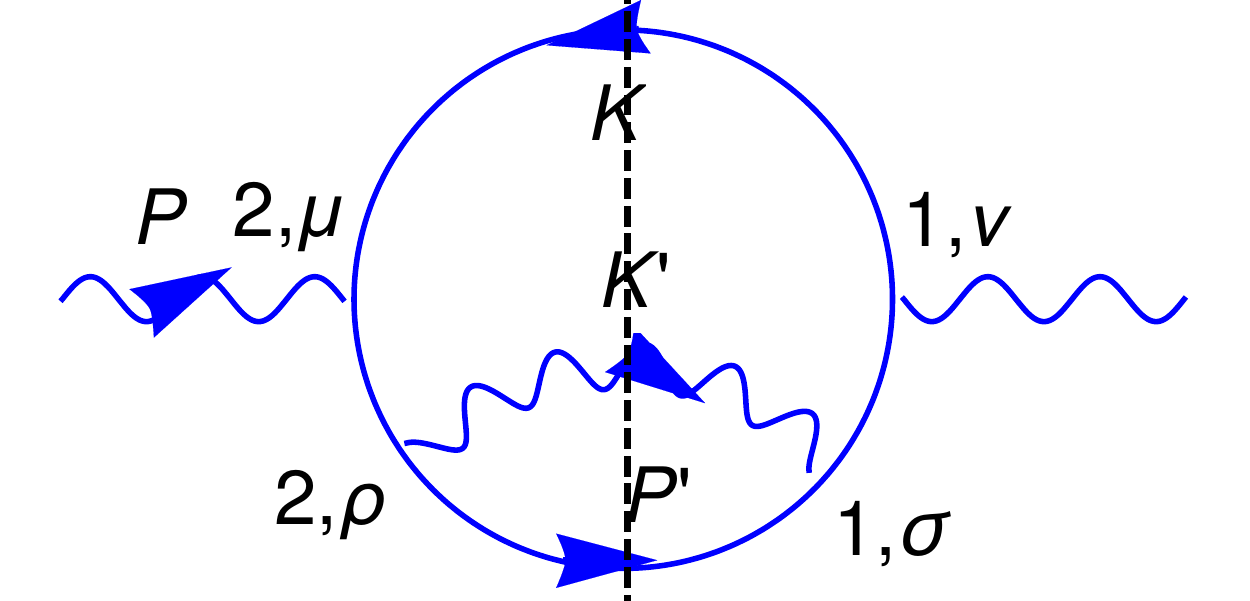}
\caption{\label{se-photon}Two-loop diagrams for photon self-energy containing propagator corrections. the $12$ labels are uniquely determined by the requirement that three propagators attached to a vertex cannot be simultaneously on-shell. The on-shell particles are indicated by a cut (dashed line) through the corresponding propagators.}
\end{figure}

Next we turn to $\P^{\m\n>}(P)D_{\n\m}^<(P)-\P^{\m\n<}(P)D_{\n\m}^>(P)$. We focus on the term $\P^{\m\n>}(P)D_{\n\r}^<(P)$. Similar to the fermionic case, we look at two-loop photon self-energy diagrams containing propagator corrections. There is only one such diagram shown in Fig.~\ref{se-photon}.
It gives the following contribution
\begin{align}\label{PD}
&\P^{\m\n>}(P)D_{\n\m}^<(P)=-e^4\int_{K,K'}\tr[\g^\n S^<(K)\g^\m S^{22}(K+P)\g^\r S^>(P')S^{11}(K+P)]\no
&\times D_{\r\s}^>(K')D_{\n\m}^<(P).
\end{align}
and a counterpart from $>\leftrightarrow<$ and $1\leftrightarrow 2$ of the above. We show in Appendix A that \eqref{PD} give rise to squares of s-channel Compton scattering, u-channel Compton scattering and t-channel annihilation.

To compare with the spin-averaged Boltzmann equations by AMY \cite{Arnold:2002zm,Arnold:2000dr,Arnold:2003zc}, we note that square of u-channel annihilation and u-channel Coulomb (between fermions) are not present in our analysis. In fact, upon integration over phase space, they give identical contributions as their t-channel counterparts. The extra contributions in Boltzmann equations are precisely taken care of by the symmetry factor $\frac{1}{2}$ for identical particles in final states applicable for annihilation and Coulomb between fermions. Therefore we find agreement on the square of amplitudes in all channels for the collision term.

\subsection{vertex corrections}

\begin{figure}
\includegraphics[width=0.45\textwidth]{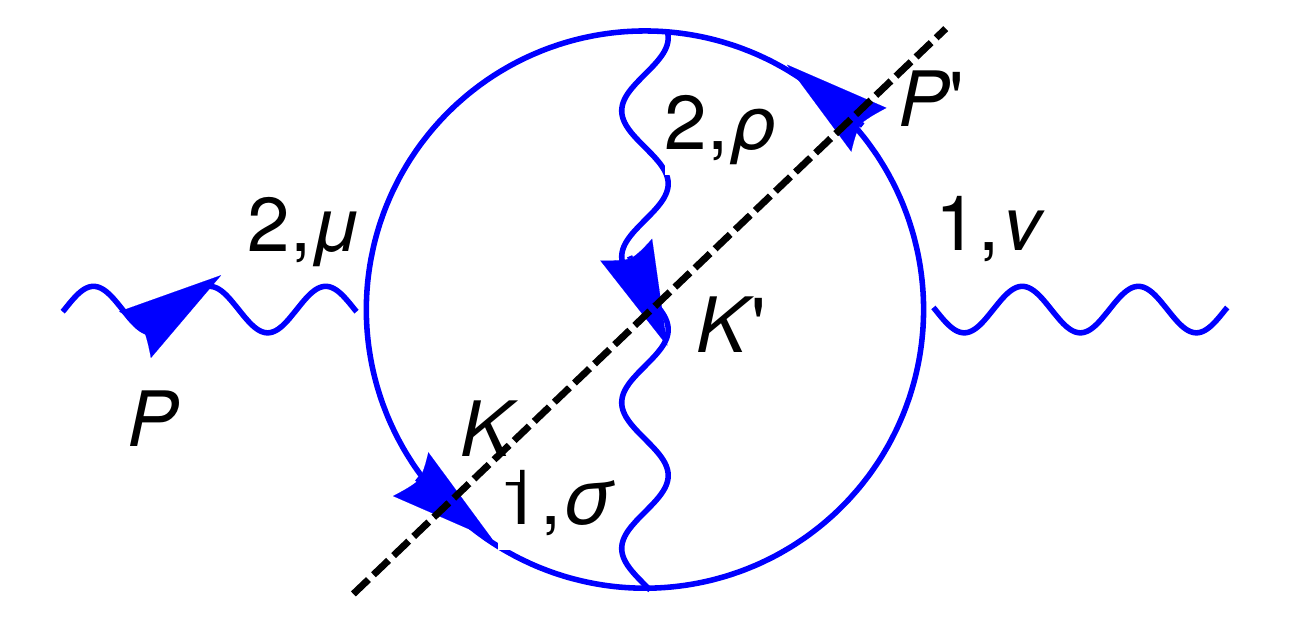}
\caption{\label{se-photon-v}Two-loop diagrams for photon self-energy containing vertex corrections. The on-shell particles are indicated by a cut (dashed line) through the corresponding propagators.}
\end{figure}

The propagator corrections give square of amplitudes only. The interference between amplitudes arise from vertex corrections, which we discuss in the following. We begin with the simpler case of vertex correction in photon self-energy. In this case, only one diagram contributes, shown in Fig.\ref{se-photon-v}.
Its contribution to collision term can be written as
\begin{align}\label{PD_v}
&\P^{\m\n>}(P)D_{\n\m}^<(P)=-e^4\int_{K,K'}\tr[\g^\m S^>(K)\g^\s S_{11}(K+K')\g^\n S^<(P')\g^\r S_{22}(K-P)]\no
&\times D_{\r\s}^>(K')D_{\n\m}^<(P),
\end{align}
and $\P^{\m\n<}(P)D_{\n\m}^>(P)$ obtainable from the above by $>\leftrightarrow <$ and $1\leftrightarrow 2$.
In Appendix A, we show this give rise to {\it half of} interference terms of s/u-channels of Compton scattering and interference term of t/u-channels of annihilation.

\begin{figure}
\includegraphics[width=0.45\textwidth]{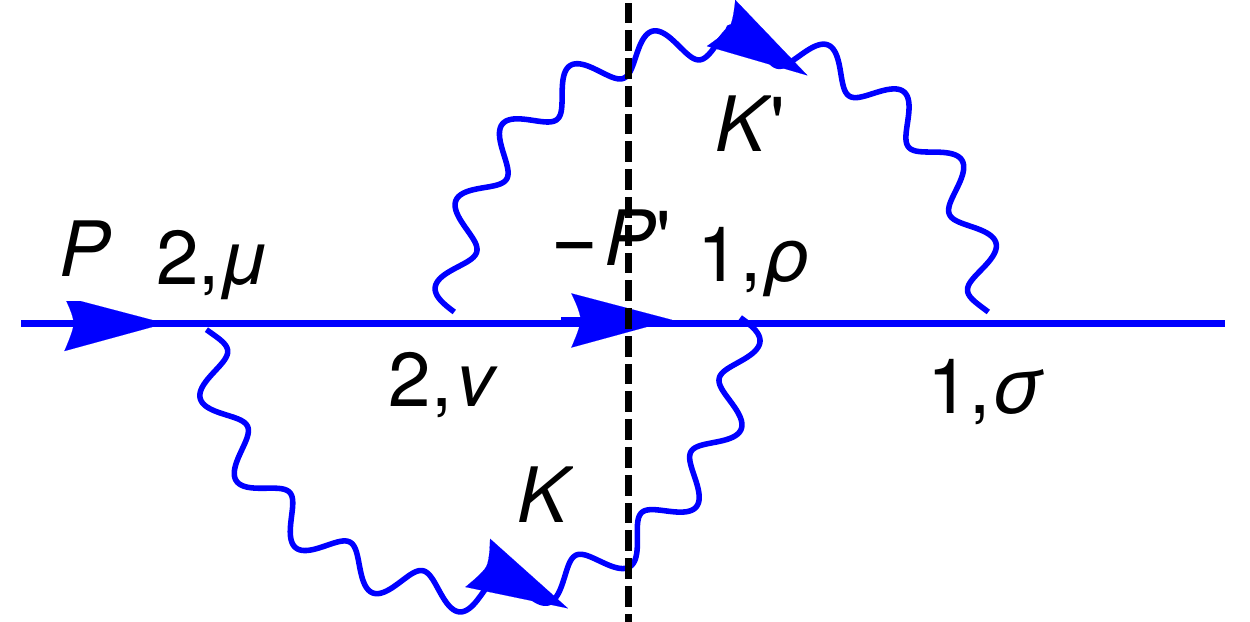}
\includegraphics[width=0.45\textwidth]{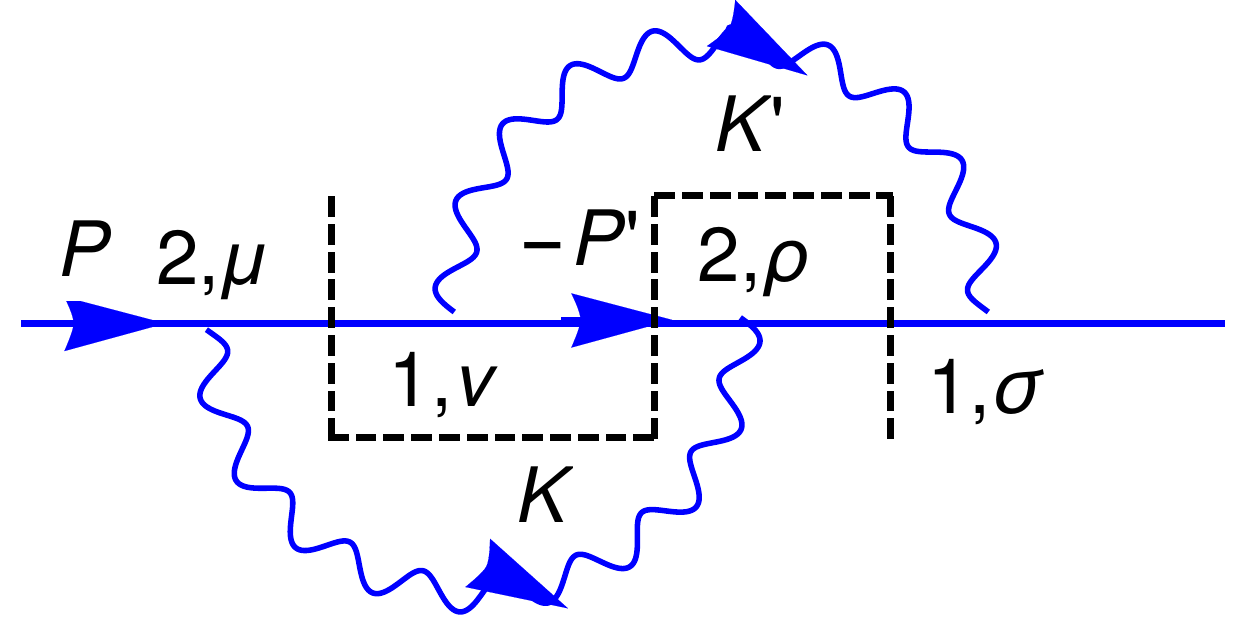}
\caption{\label{se-fermion-v}Two-loop diagrams for fermion self-energy containing vertex correction. There are two possible labelings of $12$. The on-shell particles are indicated by a cut (dashed line) through the corresponding propagators.}
\end{figure}

Now we move to the vertex correction in fermion self-energy. We have two labelings for the vertex correction diagram shown in Fig.\ref{se-fermion-v}.
For the labeling on the left, we have the following contribution to collision term
\begin{align}\label{SS_v1}
\S^>(P)S^<(P)=\int_{K,K'}\g^\m S_{22}(P-K)\g^\n S^>(-P')\g^\r S_{11}(P-K')\g^\s D_{\n\s}^>(K')D_{\m\r}^>(K)S^<(P).
\end{align}
We show in Appendix A that it gives rise to the following interference terms: {\it half of} t/u-channel for annihilation, s/t channel for Compton scattering.

For the other labeling on the right of Fig.\ref{se-fermion-v}, we have the following contribution to collision term
\begin{align}\label{SS_v2}
\S^>(P)S^<(P)=\int_{K,K'}\g^\m S^>(P-K)\g^\n S^<(-P')\g^\r S^>(P-K')\g^\s D_{\m\r}^{22}(K)D_{\n\s}^{11}(K')S^<(P).
\end{align}
It leads to the following interference terms: {\it half of} t/u-channels for Coulomb scattering (between fermions) and s/t-channels for Coulomb scattering (between fermion and anti-fermion).

Note that we have only half of the interference terms for annihilation and Coulomb scattering (between fermions). In fact in these cases, two interference terms are both real. Therefore, they again agree with the counterpart in Boltzmann equation when the symmetry factor $\frac{1}{2}$ for annihilation and Coulomb scattering (between fermions) is taken into account.

To sum up, we write down the elastic contribution to the loss terms on the RHS of \eqref{reduced_f} and \eqref{reduced_p} as
\begin{align}\label{loss_f}
&\tr[\S^{>(0)}(P)S^{<(0)}(P)]=\frac{1}{2}\int_{P',K,'K'}(2\p)^8\d^{(4)}(P+K-P'-K')\d(P^2-m^2)\frac{1}{16E_pE_{p'}E_kE_{k'}}\times\no
&\big[|{\cal M}(P,K,P',K')|_{ee\to ee}^2f_e(P)f_e(K)(1-f_e(P'))(1-f_e(K'))\d(K^2-m^2)\d(P'{}^2-m^2)\d(K'{}^2-m^2)\no
&+2|{\cal M}(P,K,P',K')|_{e\bar{e}\to e\bar{e}}^2f_e(P)f_{\bar{e}}(K)(1-f_e(P'))(1-f_{\bar{e}}(K'))\d(K^2-m^2)\d(P'{}^2-m^2)\d(K'{}^2-m^2)\no
&+|{\cal M}(P,K,P',K')|_{e\bar{e}\to\g\g}^2f_e(P)f_{\bar{e}}(K)(1+f_\g(P'))(1+f_\g(K'))\d(K^2-m^2)\d(P'{}^2)\d(K'{}^2)\no
&+2|{\cal M}(P,K,P',K')|_{e\g\to e\g}^2f_e(P)f_\g(K)(1-f_e(P'))(1+f_\g(K'))\d(K^2)\d(P'{}^2-m^2)\d(K'{}^2)],
\end{align}
and
\begin{align}\label{loss_p}
&\P^{\m\n>(0)}(P)D_{\n\m}^{<(0)}(P)=-\frac{1}{2}\int_{P',K,'K'}(2\p)^8\d^{(4)}(P+K-P'-K')\d(P^2)\frac{1}{16E_pE_{p'}E_kE_{k'}}\times\no
&\big[|{\cal M}(P,K,P',K')|_{\g\g\to e\bar{e}}^2f_\g(P)f_\g(K)(1-f_e(P'))(1-f_{\bar{e}}(K'))\d(K^2)\d(P'{}^2-m^2)\d(K'{}^2-m^2)\no
&+2|{\cal M}(P,K,P',K')|_{e\g\to e\g}^2f_\g(P)f_e(K)(1+f_\g(P'))(1-f_e(K'))\d(K^2-m^2)\d(P'{}^2)\d(K'{}^2-m^2)\no
&+2|{\cal M}(P,K,P',K')|_{\bar{e}\g\to \bar{e}\g}^2f_\g(P)f_{\bar{e}}(K)(1+f_\g(P'))(1-f_{\bar{e}}(K'))\d(K^2-m^2)\d(P'{}^2)\d(K'{}^2-m^2)],
\end{align}
respectively.
By simply exchanging the initial and final states we can obtain the corresponding gain terms, which we will not shown explicitly.

\subsection{Screening effect}

The elastic collisions occur either by exchanging off-shell photon (Coulomb scattering) or by exchanging off-shell fermion (Compton scattering and annihilation). Potential IR divergences exist when the exchange particles have soft momenta. It is known that the IR divergence can be rendered finite by the screening effect. Essentially the particles gain self-energy by interaction with the off-equilibrium medium described by the kinetic theory, which effectively cuts off the divergence. The self-energy scales as $e\L$ for both fermion and photon with $\L$ being a characteristic scale of particle energies. When $m\gg e\L$, the screening effect on Compton and annihilation is negligible: the bare mass of fermion plays the role of the cutoff. When $m\lesssim e\L$, the screening effect is non-negligible. The case of Coulomb is special. Since photon is strictly massless, the screening effect provides the only cutoff\footnote{In fact, the screening alone is not sufficient to cut off the divergence in Coulomb. A cancellation between loss and gain terms is needed to render the corresponding collision term finite.}.
We will discuss two representative scenarios: $m\gg e\L$ and $m\lesssim e\L$. For the former, we only need self-energy of photon. For the latter, we need self-energies of both fermion and photon. All the quantities studied in this subsection are local in $X$, below we suppress the dependence on $X$ for simplicity.

Now we work out the medium dependent self-energy.
We begin with the fermionic case. In elastic collisions, the exchanged off-shell fermions propagators $S_{11}/S_{22}$ can be effectively replaced by $S_R/S_A$ by using
\begin{align}\label{12RA}
S_{22}(P)=iS_A(P)+S^<(P),\quad S_{11}(P)=-iS_R(P)+S^<(P),
\end{align}
and the on-shell condition enforced by $S^{</>}$.
Note that at the lowest order in $\hb$, the retarded and advanced propagators are related by ``hermitian conjugate'' for fermions and complex conjugate for photons:
\begin{align}\label{hermiticity}
S_A(P)=\g^0S_R(P)^\dg\g^0,\quad D_{\m\n}^A(P)=D_{\n\m}^{R*}(P).
\end{align}
So it is sufficient to consider medium modification to retarded propagators only. Let us consider the fermionic retarded self-energy, which satisfies the following equation
\begin{align}\label{retarded_KE}
&i\slashed{\pd}_xS_R(x,y)-mS_R(x,y)=-\d(x-y)+\int d^4z\S_R(x,z)S_R(z,y),
\end{align}
which can be derived by taking both $x$ and $y$ in the upper branch in \eqref{hh}, and subtracting the resulting equation from \eqref{KB_raw}.
The Wigner transform of \eqref{retarded_KE} to lowest order in $\hb$ satisfies
\begin{align}\label{KB_R_f}
\(\slashed{Q}-m-\S_R(Q)\)S_R(Q)=-1,
\end{align}
from which we can solve for the resummed propagator
\begin{align}\label{fermion_R}
S_R(Q)=-\(\slashed{Q}-m-\S_R(Q)\)^{-1}.
\end{align}
$\S_R$ is evaluated in Appendix B. We quote the result here
\begin{align}\label{R_sigma}
\S_R(Q)=e^2\int \frac{d^3p}{(2\p)^3}\frac{1}{2p}\frac{\slashed{P}}{P\cdot Q}(2f_\g(\vp)+f_e(\vp)+f_{\bar{e}}(\vp)).
\end{align}
Note that this result assumes $m\lesssim e\L$ and we have dropped correction to $\S_R(Q)$ of order $e^3\L^2/Q$ from including fermion mass in the loop. This is justified because the correction is maximized at $Q\sim e\L$, for which $Q\sim m\gg e^3\L^2/Q$ so that the correction can be neglected in \eqref{KB_R_f}.

The photonic case is in parallel. The retarded propagator in Coulomb gauge satisfies
\begin{align}\label{KB_R_p}
\(-Q^2g^{\m\n}+Q^\m Q^\n-\frac{1}{\x}P^{\m\a}P^{\n\b}Q_\a Q_\b\)D_{\n\r}^R(Q)-\P^{\m\n}_R D_{\n\r}^R(Q)=-\d^\m_\r.
\end{align}
This can be solved by
\begin{align}\label{photon_R}
D_{\m\n}^R(Q)=\frac{-1}{Q^2-\P_T^R}P_{\m\n}^T+\frac{-1}{q^2+\P_L^R}u_\m u_\n+\x\frac{Q_\m Q_\n}{q^4},
\end{align}
where $\P_T^R$ and $\P_L^R$ are transverse and longitudinal components of retarded photon self-energy defined as
\begin{align}\label{se_Pi}
\P_{\m\n}^R=P_{\m\n}^T\P_T^R-\frac{Q^2}{q^2}P_{\m\n}^L\P_L^R.
\end{align}
We have introduced the longitudinal projector $P_{\m\n}^L=-g_{\m\n}+\frac{Q_\m Q_\n}{Q^2}-P_{\m\n}^T$ and $q^2=-Q^2+(Q\cdot u)^2$. We again set $\x=0$ for Coulomb gauge. $\P_{\m\n}^R$ is evaluated in Appendix B. We quote the result here\footnote{For an anisotropic medium, for example medium with shear gradient, gradient correction to $f_{e/\bar{e}}$ can give rise to additional structures not present in \eqref{se_Pi}. These corrections introduce a subleading effect on screening, which is suppressed by the factor $\frac{\pd_Xf}{\L}\ll1$. We can ignore such correction in screening as long as we are concerned with leading order effect of gradient.}
\begin{align}\label{R_Pi}
\P_{\m\n}^R(Q)=2e^2\int\frac{d^3p}{(2\p)^3}\frac{1}{E_p}(f_e(\vp)+f_{\bar{e}}(\vp))\big[\frac{P_\m Q_\n+P_\n Q_\m-g_{\m\n}P\cdot Q}{P\cdot Q}-\frac{P^\m P^\n Q^2}{(P\cdot Q)^2}\big],
\end{align}
which is applicable for both scenarios with mass dependence implicit in $E_p=\sqrt{p^2+m^2}$. We iterate that the following replacement with resummed propagators are to be used in the evaluation of self-energies when exchanged particles have soft momenta:
\begin{align}
&S_{22}(P)\to iS_A(P),\quad S_{11}(P)\to -iS_R(P),\no
&D^{\m\n}_{22}(P)\to iD^{\m\n}_A(P),\quad D^{\m\n}_{11}(P)\to -iD^{\m\n}_R(P).
\end{align}

We will also need the asymptotic thermal mass for hard on-shell fermions and transverse photons in inelastic collisions in the next section, which we determine below. \eqref{KB_R_f} determines the fermion dispersion relation by:
\begin{align}\label{disp_matrix}
\slashed{Q}-m-\S_R(Q)=0.
\end{align}
Following \cite{Petitgirard:1991mf}, we decompose the self-energy as $\S_R=\S^R_\m\g^\m+\S^R_m1$, which allows us to convert the matrix equation \eqref{disp_matrix} into a scalar equation
\begin{align}\label{disp}
(Q_\m-\S^R_\m)^2-(\S^R_m+m)^2=0.
\end{align}
From \eqref{R_sigma}, we have $\S^R_\m\sim e^2\L$ and $\S^R_m=0$. The thermal mass is identified as
\begin{align}\label{mth_e}
\d m_e^2=2 Q^\m\S^R_\m=e^2\int\frac{d^3p}{(2\p)^3}\frac{2}{p}\(2f_\g(\vp)+f_e(\vp)+f_{\bar{e}}(\vp)\).
\end{align}
The only change from the resummed propagator in equilibrium is that the thermal mass depends on the off-equilibrium distributions for fermion and photon.

The photon thermal mass is determined from the pole of the transverse component of propagator as:
\begin{align}
m_\g^2=\P_T^R=\frac{1}{2}P_{\m\n}^T\P_R^{\m\n}.
\end{align}
The term $P_\m Q_\n+P_\n Q_\m$ in \eqref{R_Pi} is eliminated upon contraction with $P_{\m\n}^T$ and the term containing $Q^2$ vanishes by the on-shell condition, leaving the following $Q$-independent expression
\begin{align}
m_\g^2=e^2\int\frac{d^3p}{(2\p)^3}\frac{2}{E_p}(f_e(\vp)+f_{\bar{e}}(\vp)).
\end{align}

Summarizing this section, by considering propagator and vertex corrections, we have captured the full $2\to2$ elastic collision including screening effect in Boltzmann equation for QED.

\section{Inelastic collisions}\label{sec_inec}

\begin{figure}
\includegraphics[width=0.45\textwidth]{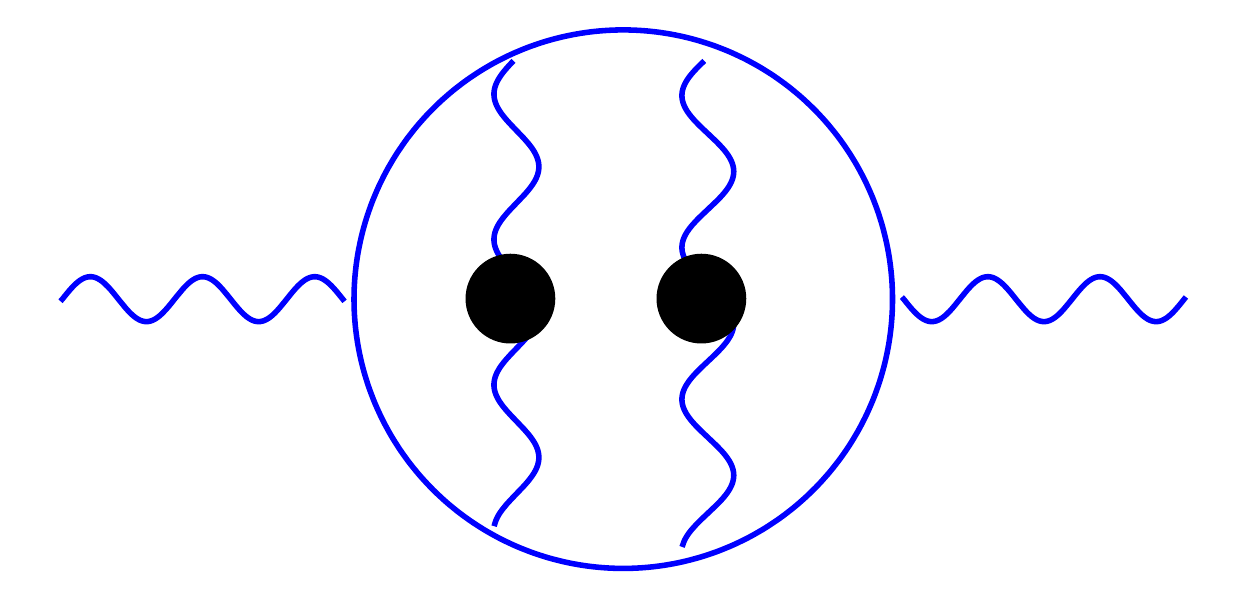}
\caption{\label{se-pinch}Photon self-energy containing vertex corrections from multiple scatterings. Arbitrary number of soft photon exchanges is possible. For illustration purpose we show two photon exchanges. The black dots indicate resummed photon propagators.}
\end{figure}

In the analysis of elastic scattering, we have excluded one-loop contribution to self-energies, which requires all three particles to be exactly collinear, leaving vanishing phase space. In fact, it is not entirey excluded: when we take into account medium modification of energy of on-shell degree of freedoms, the energy conservation can be slightly violated, opening up a small phase space.
It is known that in equilibrium medium both fermion and photon gain thermal mass and damping rate, which modifes the energy by an amount of order $e^2\L$. Such a modification can also be realized by transverse momenta of order $e\L$. It implies that medium modification to energy allows for deviation of collinearity with transverse momenta of order $e\L$, which gives rise to a phase space $\frac{d^2p_\pp dp_{\pr}}{E_p}\sim e^2\L^2$. Combining this with $e^2$ from vertices in one-loop diagram, we have an overall $e^4\L^2$, which is the same as the two-loop diagrams.
In fact, fermions can have multiple soft scatterings with the medium, which by pinching mechanism all contribute at the same order and need to be added up coherently, known as Landau-Pomeranchuk-Migdal effect. The multiple scatterings are encoded in another type of vertex corrections in self-energies diagrams, shown in Fig.~\ref{se-pinch}. It involves multiple insertions of soft photon propagators. When there is no insertion of photon propagator, it reduces to a one-loop diagram, for which the power counting is done above. Below we confirm that the power counting is not affected by insertions of photon progagators.

\subsection{Resummed vertex}

\begin{figure}
\includegraphics[width=0.8\textwidth]{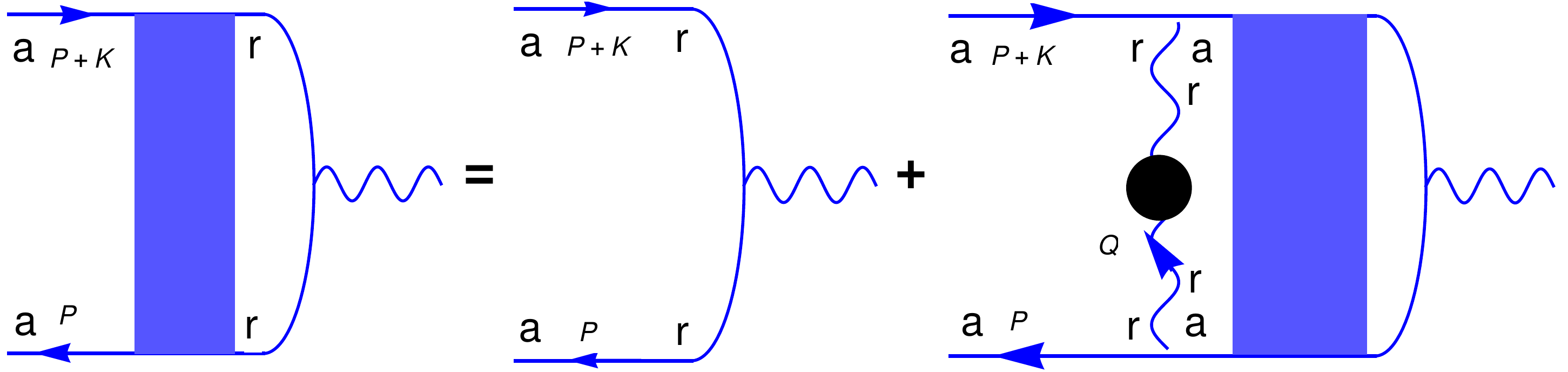}
\caption{\label{diag-eq}Diagrammatic equation satisfied by the resummed vertex. We will consider spin-dependent vertex, in which the fermions with momenta $P+K$ and $P$ carrying spin labels $s$ and $t$.}
\end{figure}

Let us illustrate the pinching mechanism by looking at the resummed vertex. Following \cite{Arnold:2001ba}, the evaluation is most conveniently done in the $ra$ basis, in which only one inequivalent labeling is allowed\footnote{The other one is related by interchanging $r$ and $a$. All other labelings can be related to the two.}. The resummed vertex satisfies the diagrammatic equation in Fig.~\ref{diag-eq}.
The pinching mechanism can be seen by inspecting the energy integration of two side-rails
\begin{align}\label{side-rail}
\int\frac{dp_0}{2\p}S_{ra}(P)(\cdots)S_{ar}(K+P),
\end{align}
with $S_{ra}=D_{ra}(P)(\slashed{P}+m)$ and $S_{ar}=D_{ar}(K+P)(\slashed{K}+\slashed{P}+m)$. $D_{ra}$ and $D_{ar}$ are the same as scalar propagators given by\footnote{$\frac{\G}{2}$ is defined as the damping rate here.}
\begin{align}\label{D_raar}
&D_{ra}(P)=\frac{i}{(p_0+\frac{i}{2}\G_p)^2-E_p^2},\no
&D_{ar}(K+P)=\frac{i}{(p_0+k_0-\frac{i}{2}\G_{p+k})^2-E_{p+k}^2}.
\end{align}
To be specific, we take $k_0>0$, Let us evaluate the following by residue theorem
\begin{align}
\int\frac{dp_0}{2\p}D_{ra}(P)D_{ar}(K+P).
\end{align}
The poles are located at $p_0=\pm E_p-\frac{i}{2}\G_p$, $p_0=-k_0\pm E_{p+k}+\frac{i}{2}\G_{p+k}$. Here $P$ is allowed to have transverse component $p_\pp\sim e\L$. Here the energies of both fermions and photon receive corrections from deviation of collinearity and thermal/bare mass. Note that the photon momentum is used to define longitudinal direction, so it has no deviation from collinearity.
\begin{align}\label{fp_disp}
&E_p=p+\frac{p_\pp^2+\d m_e^2+m^2}{2p},\no
&E_{p+k}=|{\bf p}+{\bf k}|+\frac{p_\pp^2+\d m_e^2+m^2}{2|{\bf p}+{\bf k}|},\no
&k_0=k+\frac{m_\g^2}{2k},
\end{align}
Denoting $p_\pr=\vp\cdot\hat{k}$, we can rewrite the poles as
$p_0=\pm p_\pr-\frac{i}{2}\G_p$, $p_0=-k_0\pm (p_\pr+k)+\frac{i}{2}\G_{p+k}$.
The pinching mechanism is at work when two of the poles nearly pinch. 
Ignoring the thermal masses and damping rate, we find the poles at $p_0=p_\pr$ and $p_0=-k_0+p_\pr+k$ coincide. The thermal masses and damping rates provides necessary regularization to the divergence. Closing the contour and picking up one of the pinching poles, we obtain
\begin{align}\label{pinch_integral}
&\int\frac{dp_0}{2\p}D_{ra}(P)D_{ar}(K+P)
=\frac{-1}{4p_\pr(p_\pr+k)(i\d E+\G)},
\end{align}
with
\begin{align}
&\d E=k_0+E_p\e(p_\pr)-E_{p+k}\e(p_\pr+k)\simeq\frac{k(p_\pp^2+\d m_e^2+m^2)}{2p_\pr(p_\pr+k)}+\frac{m_\g^2}{2k},\no
&\G=\frac{1}{2}\(\G_p+\G_{p+k}\).
\end{align}
In addition, each pair of pinching propagators (side-rails) is accompanied by a soft photon propagator (rung). The power counting for the latter is done as follows: note that the momentum $Q$ of the off-shell photon is spacelike and its time component is fixed the pinching conditions as $\int dq_0\d(p_0-p_\pr)\d(p_0-q_0-p_\pr-q_\pr)=\d(p_0-p_\pr)$, so that the phase space for $Q$ counts as $e^3$. The power counting for the relevant propagator $D_{rr}(Q)$ can be done using its equilibrium form $D_{rr}(Q)=(\frac{1}{2}+f(q_0))\r(Q)$. Using that the spectral density $\r(Q)\sim Q^{-2}\sim e^{-2}$ and the Bose enhancement factor $f(q_0)\sim q_0^{-1}\sim e^{-1}$, we have $D_{rr}(Q)\sim e^{-3}$. Combining pieces together with an $e^2$ from extra vertices, we obtain for each one pair of side-rails and one rung
\begin{align}\label{pinch_power}
\frac{1}{i\d E+\G}e^3e^{-3}e^2\sim\frac{e^2}{i\d E+\G}
\end{align}

In the scenario $m\lesssim e\L$, $\d E\sim e^2\L$ and $\G$ is evaluated in appendix C to be at the same order. Thus we have from \eqref{pinch_power} an overall factor $e^0$, thus arbitrary number of insertions are allowed, corresponding to multiple soft scatterings of hard fermions with the medium fermions.
However in the scenario $m\gg e\L$, the pinching mechanism is suppressed by a factor of $\frac{e^2\L^2}{m^2}$. It follows that inelastic scattering is irrelevant in this scenario to the order of our interest. Below we proceed with the scenario $m\lesssim e\L$.

We denote the resummed vertex in Fig.~\ref{diag-eq} as $\G_{st}^\m(P+K,P)$, with $s,t$ labeling the spinors in the outermost propagators carrying momenta $P+K$ and $P$ respectively. This is a polarization-dependent vertex between two on-shell fermions with spin $s,t$ and one on-shell photon. We can take the Lorentz index $\m$ to be transverse, since the resulting self-energy will be contracted with projector in \eqref{reduced_p}. Below we denote $\m$ by $M$ for transverse indices. Below we first consider the case $p_0>0$, for which $\G_{st}^M$ can be interpreted as the amplitude of a fermion (with momentum $P+K$ and spin $s$) splitting into a fermion (with momentum $P$ and spin $t$) and a photon (with momentum $K$ and polarization $\varepsilon^M$).
We can express the diagrammatic equation as\footnote{An overall $-ie$ is factored out from the vertex.}
\begin{align}\label{diag_int_eq}
&\G_{st}^M(P+K,P)=D_{ra}(P)D_{ar}(P+K)u_s(P+K)\bar{u}_s(P+K)\g^M u_t(P)\bar{u}_t(P)\no
&+\int_QD_{ra}(P)D_{ar}(P+K)\((\slashed{P}+\slashed{K})(-ie\g^\m)\G_{st}^M(P+K+Q,P+Q)(-ie\g^\n)\slashed{P}\)D_{\n\m}^{rr}(Q),
\end{align}
where the structures $u_s(P+K)\bar{u}_s(P+K)$ and $u_t(P)\bar{u}_t(P)$ come from $\slashed{P}+\slashed{K}$ and $\slashed{P}$ projected onto given spin states respectively.
Note that we have dropped the subleading masses in the numerators of fermion propagators. We will use the following representation for spinors
\begin{align}
u_s(P)=\sqrt{\frac{p_0}{2}}
\begin{pmatrix}
(1-\s\cdot\vp/p_0)\x_s\\
(1+\s\cdot\vp/p_0)\x_s
\end{pmatrix},
\end{align}
with $\x_+=(1,0)^T$ and $\x_-=(0,1)^T$. In this representation, the spin label $s=\pm$ corresponds to spin being parallel/anti-parallel to momentum. We can derive the following relations valid to leading order in $e$
\begin{align}\label{spin_contractions}
&\bar{u}_s(P+K)\g^\m u_{s'}(P+Q+K)\simeq \d_{ss'}2(P+K)^\m,\no
&\bar{u}_{t'}(P+Q)\g^\n u_{t}(P)\simeq \d_{tt'}2P^\n,\no
&\bar{u}_s(P+K)\g^M u_{t}(P)\simeq \(p_\pr(p_\pr+k)\)^{-1/2}\((p_\pr+k)p_\pr^M+p_\pr p_{-s}^M\)\d_{st},
\end{align}
with $p_{\pm}^M=p^M\pm i\e^{MN}p^N$ and we have used the pinching condition $p_0\simeq p_\pr$. We easily deduce the spin direction of the fermion along the upper/lower rails is not changed. We can parametrize the vertex by
\begin{align}\label{vertex_para}
\G_{st}^M(P+K,P)=\d_{st}u_s(P+K)\bar{u}_t(P)\(p_\pr(p_\pr+k)\)^{-1/2}\((p_\pr+k)p_s^M+p_\pr p_{-s}^M\)\G_s(P).
\end{align}
By rotational invariance, $\G(P)$ is a function of $p_\pr$, $p_\pp^2$. It follows then
\begin{align}\label{Gs_eq}
&\((p_\pr+k)p_s^M+p_\pr p_{-s}^M\)\G_s(P)=\((p_\pr+k)p_s^M+p_\pr p_{-s}^M\)D_{ra}(P)D_{ar}(P+K)\no
&-e^2\int_QD_{ra}(P)D_{ar}(P+K)\((p_\pr+k)(p+q)_s^M+p_\pr(p+q)_{-s}^M\)\G_s(P+Q)4p_\pr(p_\pr+k)\no
&\times\hat{K}^\m\hat{K}^\n D_{\n\m}^{rr}(Q).
\end{align}
By pinching mechanism, the vertex should have support localized on the pole $p_0=p_\pr$. The same kinematic restriction should apply to the bare vertex in the diagrammatic equation in Fig.~\ref{diag-eq}. It is then convenient to define
\begin{align}
\int\frac{dp_0}{2\p}\G_s(P)=-\frac{1}{4p_\pr(p_\pr+k)}\c_s(P).
\end{align}
In terms of $\c_s$, \eqref{Gs_eq} becomes
\begin{align}\label{Cs_eq}
&-\((p_\pr+k)p_s^M+p_\pr p_{-s}^M\)\c_s(P)(i\d E+\G)=-\((p_\pr+k)p_s^M+p_\pr p_{-s}^M\)\no
&-e^2\int\frac{d^3q}{(2\p)^3}\((p_\pr+k)(p+q)_s^M+p_\pr(p+q)_{-s}^M\)\c_s(P+Q)\hat{K}^\m\hat{K}^\n D_{\n\m}^{rr}(Q).
\end{align}

\eqref{Cs_eq} can be further simplified by using the following representation of $\G$:
\begin{align}\label{G_rep}
\G=e^2\int\frac{d^3q}{(2\p)^3}\hat{K}^\m\hat{K}^\n D_{\n\m}^{rr}(Q).
\end{align}
It contains the same soft photon propagator as in \eqref{Cs_eq}. A derivation of the representation can be found in Appendix C. Using \eqref{G_rep}, we can rewrite \eqref{Cs_eq} as
\begin{align}\label{int_eq}
&\((p_\pr+k)p_s^M+p_\pr p_{-s}^M\)=\((p_\pr+k)p_s^M+p_\pr p_{-s}^M\)\c_s(P)i\d E+e^2\int\frac{d^3q}{(2\p)^3}\hat{K}^\m\hat{K}^\n D_{\n\m}^{rr}(Q)\no
&\big[\((p_\pr+k)p_s^M+p_\pr p_{-s}^M\)\c_s(P+Q)-\((p_\pr+k)(p+q)_s^M+p_\pr(p+q)_{-s}^M\)\c_s(P+Q)\big].
\end{align}

The structure $(p_\pr+k)p_s^M+p_\pr p_{-s}^M$ encodes the spin dependence of the vertex. It is more transparent to switch to circular polarizations for the photon $\varepsilon_\pm^{M}=\frac{1}{\sqrt{2}}(0,1,\pm i,0)$. The corresponding coordinates and momenta are defined by $x^{\pm}=\frac{1}{\sqrt{2}}(x\pm iy)$ and $p^{\pm}=\frac{1}{\sqrt{2}}(p_x\pm ip_y)$, with $M=\pm$ in the circular basis.
Using $p_+^M=(\sqrt{2}p^+,0)$, $p_-^M=(0,\sqrt{2}p^-)$, we find \eqref{int_eq} splits into two cases $s=\pm M$, which satisfy a unified equation
\begin{align}\label{unified_eq}
p^M=p^M\c_s(P)i\d E+e^2\int\frac{d^3q}{(2\p)^3}\hat{K}^\m\hat{K}^\n D_{\n\m}^{rr}(Q)
\big[p^M\c_s(P+Q)-(p+q)^M\c_s(P+Q)\big].
\end{align}
Recall in our $\hb$ expansion, $f_e$ is spin independent. It follows from \eqref{Pi_aa_final} and \eqref{Drr} that the kernel $\hat{K}^\m\hat{K}^\n D_{\n\m}^{rr}(Q)$ is also spin independent. A spin independent $\c_s=\c$ is expected from \eqref{unified_eq}. Furthermore, \eqref{unified_eq} is manifestly rotational invariant, which is equally valid in the orthogonal basis. \eqref{unified_eq} in the orthogonal basis is in agreement with Eqs.(2.2) and (2.6) of \cite{Arnold:2002zm}, with the identification $\c=\frac{1}{2}\c_{AMY}$.

Although spin information is averaged out in the resulting kinetic theory, the spin dependence of the vertex is instructive on its own. The structure $(p_\pr+k)p_s^M+p_\pr p_{-s}^M$ in circular basis indicates that a fermion with spin in either direction can radiate a right/left handed photon without changing its spin. Clearly spin angular momentum is not conserved. Since collision is local, orbital angular momentum is zero before and after the collision. The change of spin angular momentum comes from spin exchange between the fermion/photon in the resummed vertex and fermions in the medium, whose spin information is averaged out.



\subsection{photon self-energy}

From Fig.~\ref{diag-eq}, the resummed vertex $\G_{st}^M$ contains the four-fermi correlators $G_{aarr}(P+K,P,P+K+Q,P+Q)$ with $P+K$ and $P$ labeled by $a$ and $P+K+Q$ and $P+Q$ labeled by $r$ as in Fig.~\ref{diag-eq}. In the pinching kinematical region, $Q\ll P\&K$. To convert to photon self-energy, we need $G_{1122}(P+K,P,P+K+Q,P+Q)$. The conversion involves an off-equilibrium generalization of Kubo-Martin-Schwinger (KMS) relation. We give a diagrammatic derivation of the relation in the pinching kinematical region in Appendix D. The off-equilibrium relation simply replaces the equilibrium distribution by the off-equilibrium counterpart as
\begin{align}\label{KMS_off}
&G_{1122}(P+K,P,P+K+Q,P+Q)=\no
&f_e(p_\pr+k)(1-f_e(p_\pr))2\text{Re}G_{aarr}(P+K,P,P+K+Q,P+Q).
\end{align}
Accordingly, to obtain the photon self-energy $\P^<(K)=\int_{P,Q} G_{1122}(P+K,P,P+K+Q,P+Q)$, we should use $f_e(p_\pr+k)(1-f_e(p_\pr))2\text{Re}[\G_{st}^M]$ as the correct resummed vertex. Now we can trace $-ie\G_{st}^M(P+K,P)$ with $-ie\g^N$, include $-1$ from fermion loop, sum over spins and integrate over momentum to obtain
\begin{align}\label{se_p}
&\P^{<MN}(K)=\int_P(-1)(-ie)^2\sum_{s,t}\tr[\G_{st}^M(P+K,P)\g^N] \no
&=e^2\sum_s\int\frac{d^3p}{(2\p)^3}\frac{1}{4\(p_\pr(p_\pr+k)\)^{3/2}}\((p_\pr+k)p_s^M+p_\pr p_{-s}^M\)f_e(p_\pr+k)(1-f_e(p_\pr))\no
&\times 2\text{Re}[\c(P)]\tr[u_s(P+K)\bar{u}_s(P)\g^N].
\end{align}
Recall in \eqref{reduced_p}, $\P^{<MN}$ is contracted with $D^>_{MN}(K)\propto P_{MN}^T f_\g(k)$. We can use the following for the trace
\begin{align}
\tr[u_s(P+K)\bar{u}_s(P)\g^N]=\(p_\pr(p_\pr+k)\)^{-1/2}\((p_\pr+k)p_{-s}^N+p_\pr p_s^N\).
\end{align}
By the $s$-independent identity
\begin{align}
\sum_Mp_s^Mp_{-s}^M=2p_\pp^2,\quad \sum_Mp_s^Mp_s^M+p_{-s}^Mp_{-s}^M=0,
\end{align}
we can see the spin sum simply gives a factor of $2$. In the end, we arrive at
\begin{align}\label{coll_p}
\P^{<MN}D^>_{MN}=e^2\int\frac{d^3p}{(2\p)^3}f_e(p_\pr+k)(1-f_e(p_\pr))(1+f_\g(k))2\text{Re}[\c(P)]\frac{\((p_\pr+k)^2+p_\pr^2\)p_\pp^2}{\(p_\pr(p_\pr+k)\)^2}.
\end{align}
The other collision term $\P^{>MN}D^<_{MN}$ is obtainable from \eqref{coll_p} by the replacement of the distribution functions $f_e(p_\pr+k)(1-f_e(p_\pr))(1+f_\g(k))\to (1-f_e(p_\pr+k))f_e(p_\pr)f_\g(k)$. 

So far, we have focused on the kinematical region with $p_0>0$ and $k_0>0$. Let us first lift $p_0>0$. Other possible kinematical regions are $p_0+k<0$ corresponding to anti-fermion bremsstrahlung and $-k<p_0<0$ corresponding to inelastic annihilation. For $p_0+k<0$, we use the following projections for spin states.
\begin{align}
&\slashed{P}+\slashed{K}\to-u_s(-P-K)\bar{u}_s(-P-K),\no
&\slashed{P}\to-u_t(-P)\bar{u}_t(-P).
\end{align}
The contraction of vertices is modified slightly from \eqref{spin_contractions} as
\begin{align}
&\bar{u}_s(-P-K)\g^\m u_{s'}(-P-Q-K)\simeq -\d_{ss'}2(P+K)^\m,\no
&\bar{u}_{t'}(-P)\g^\n u_{t}(-P-Q)\simeq -\d_{tt'}2P^\n,\no
&\bar{u}_s(-P-K)\g^M u_{t}(-P)\simeq \(p_\pr(p_\pr+k)\)^{-1/2}\((p_\pr+k)p_\pr^M+p_\pr p_{-s}^M\)\d_{st}.
\end{align}
With these, we easily confirm \eqref{coll_p} is also applicable for $p_0<-k$. The analysis for $-k<p_0<0$ is similar, with $p_\pr(p_\pr+k)$ replaced by $-p_\pr(p_\pr+k)$, which leaves \eqref{coll_p} unchanged. This justifies extending the integration domain of $p_\pr$ to $(-\infty,\infty)$.

Finally we comment on the case $k_0<0$. This case would be needed for the dynamics of photon's anti-particle. Since photon's anti-particle is itself, we expect it to give an equivalent equation.

\subsection{fermion self-energy}

\begin{figure}
\includegraphics[width=0.3\textwidth]{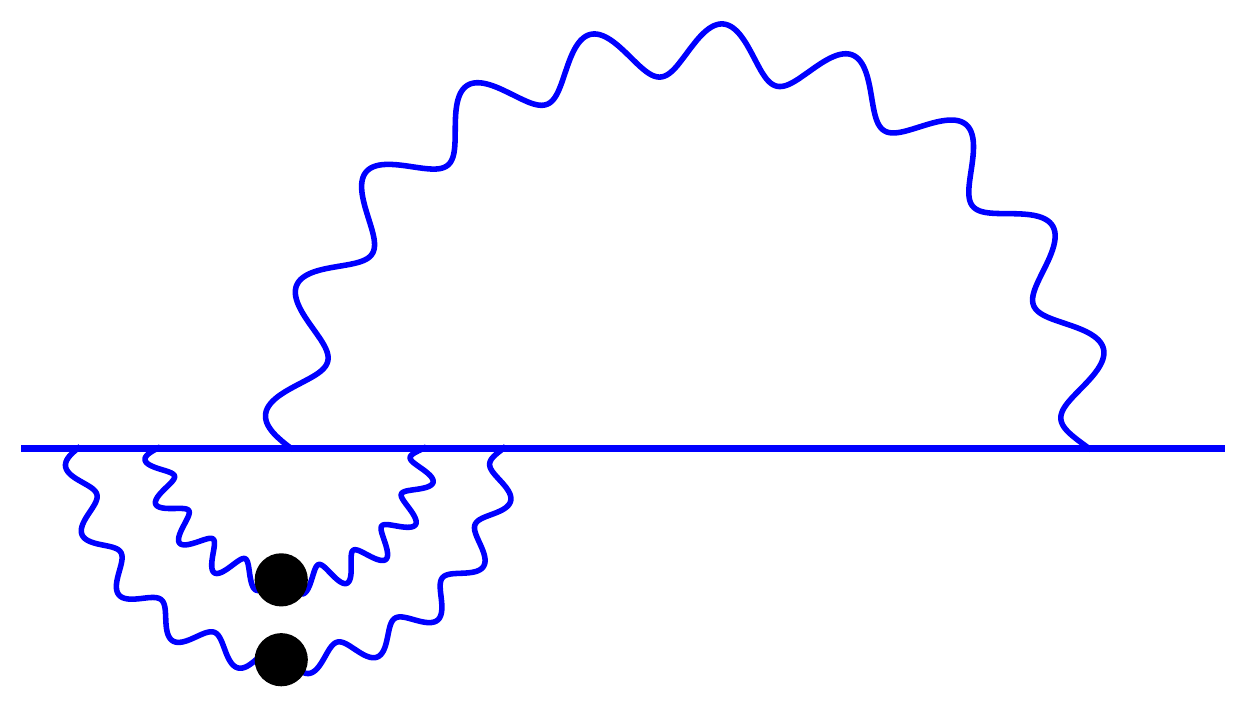}
\includegraphics[width=0.3\textwidth]{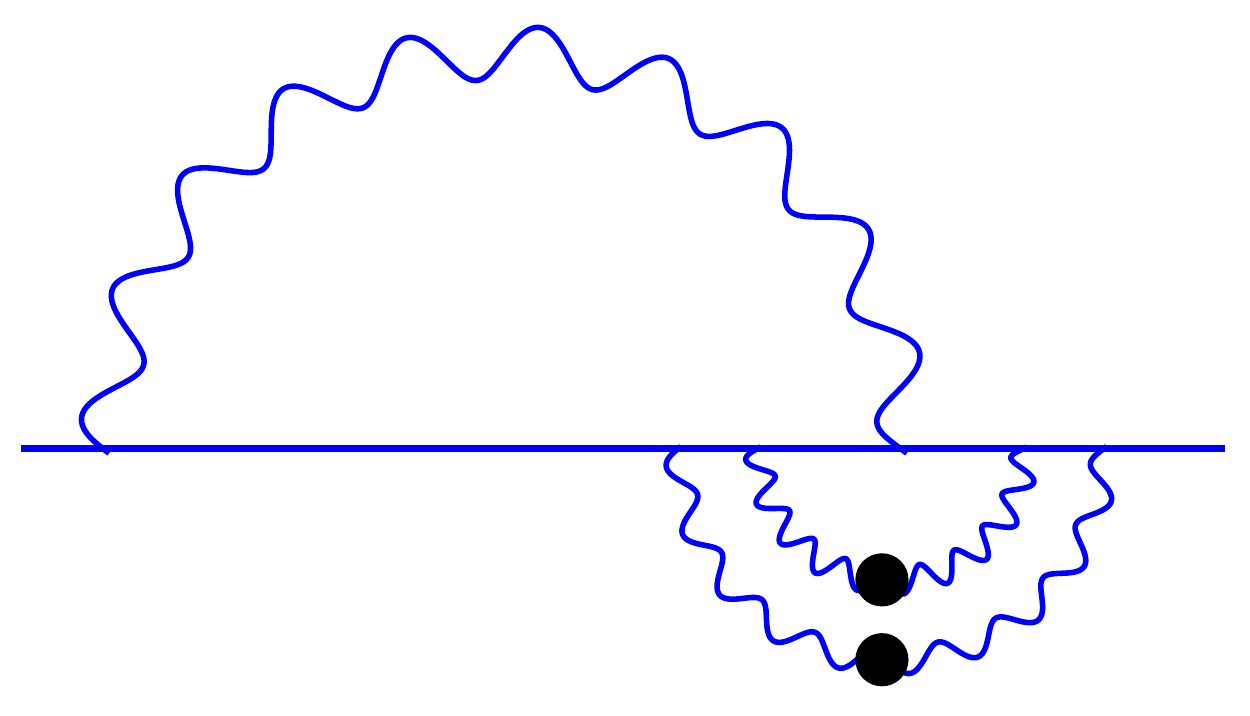}
\caption{\label{se-pinch2}Fermion self-energy containing vertex corrections from multiple scatterings. Arbitrary number of soft photon exchanges is possible. Unlike the photonic case, there are two inequivalent vertex corrections.}
\end{figure}

\begin{figure}
\includegraphics[width=0.8\textwidth]{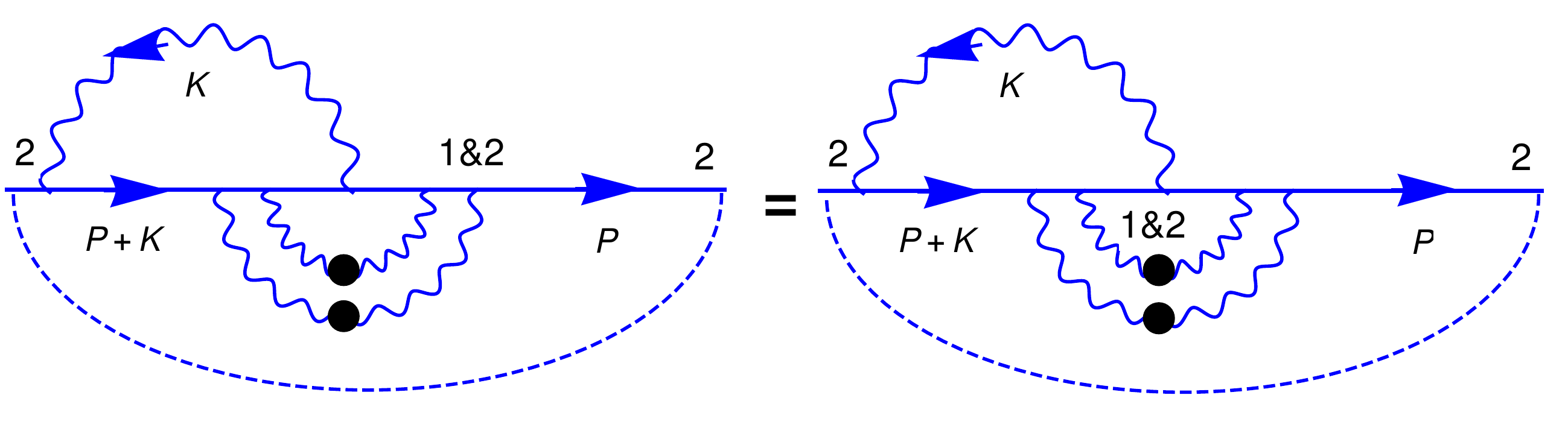}
\caption{\label{trace-eq}The LHS and RHS has one-to-one correspondence with the first line of \eqref{SS-PD}. The symbol $1\&2$ indicates the vertex can be labeled by either $1$ or $2$, giving rise to two terms. The dashed line corresponds to the trace in the left diagram and the contraction in the right diagram. Replacing the outermost labels $2$ by $1$ leads to the second line of \eqref{SS-PD}.}
\end{figure}

Unlike the photonic case, the fermion self-energy arises from two diagrams shown in Fig.\ref{se-pinch2}.
To be specific, let us consider $p_0>0$ corresponding to fermion. We consider the term $\S^>S^<(\S^<S^>)$ and focus on the second diagram first. It is convenient to include the contribution $\S_{22}S_{22}(\S_{11}S_{11})$. Upon taking the trace, we have from the diagrammatic identity in Fig.~\ref{trace-eq} that
\begin{align}\label{SS-PD}
&\tr[\S^>(P)S^<(P)+\S_{22}(P)S_{22}(P)]=\int_K[G^{\m\n>}(P+K,P)D_{\n\m}^<(K)+G_{22}^{\m\n}(P+K,P)D^{22}_{\n\m}(K)],\no
&\tr[\S^<(P)S^>(P)+\S_{11}(P)S_{11}(P)]=\int_K[G^{\m\n<}(P+K,P)D_{\n\m}^<(K)+G_{11}^{\m\n}(P+K,P)D^{11}_{\n\m}(K)],
\end{align}
where $G^{\m\n}(P+K,P)$ are partially-integrated photon self-energy defined as
\begin{align}
&G^{\m\n>}(P+K,P)=\int_Q\tr[(-ie\g^\m)G_{2211}(P+K,P,P+K+Q,P+Q)(-ie\g^\n)],\no
&G^{\m\n<}(P+K,P)=\int_Q\tr[(-ie\g^\m)G_{1122}(P+K,P,P+K+Q,P+Q)(-ie\g^\n)],\no
&G^{\m\n}_{22}(P+K,P)=\int_Q\tr[(-ie\g^\m)G_{2222}(P+K,P,P+K+Q,P+Q)(-ie\g^\n)],\no
&G^{\m\n}_{11}(P+K,P)=\int_Q\tr[(-ie\g^\m)G_{1111}(P+K,P,P+K+Q,P+Q)(-ie\g^\n)].\nonumber
\end{align}
Taking the difference of the two lines in \eqref{SS-PD}, we obtain (with arguments suppressed)
\begin{align}\label{SS_extended}
&\tr[\S^>S^<-\S^<S^>+\S_{22}S_{22}-\S_{11}S_{11}]\no
&=\int_K[G^{\m\n>}D_{\n\m}^>-G^{\m\n<}D_{\n\m}^<+G^{\m\n}_{22}D_{\n\m}^{22}-G^{\m\n}_{11}D_{\n\m}^{11}].
\end{align}
Using that $\S^{>/<}$ and $S^{>/<}$ are ``hermitian'', we can easily show $\tr[\S^>S^<-\S^<S^>]$ is real. Further using the following hermitian properties
\begin{align}\label{hermitian}
&\S_{22}=\g^0\S_{11}^\dg\g^0,\quad S_{22}=\g^0S_{11}^\dg\g^0,\no
&G^{\m\n}_{22}=G_{11}^{\n\m*},\quad D^{\m\n}_{22}=D_{11}^{\n\m*},
\end{align}
we obtain from \eqref{SS_extended}
\begin{align}\label{SS_2nd}
&\tr[\S^>(P)S^<(P)-\S^<(P)S^>(P)]=\text{Re}\(\tr[\S^>(P)S^<(P)-\S^<(P)S^>(P)]\)\no
=&\int_K[G^{\m\n>}(P+K,P)D_{\n\m}^<(K)-G^{\m\n<}(P+K,P)D_{\n\m}^>(K)],
\end{align}

The evaluation of the first diagram proceeds similarly. We have instead
\begin{align}\label{SS_1st}
&\tr[\S^>(P)S^<(P)-\S^<(P)S^>(P)]\no
=&\int_K[G^{\m\n<}(P,P+K)D_{\n\m}^<(K)-G^{\m\n>}(P,P+K)D_{\n\m}^>(K)].
\end{align}
Note that the momenta argument in \eqref{SS_2nd} and \eqref{SS_1st} differ. Using $G^{\m\n<}(P,P+K)=G^{\n\m>}(P+K,P)$ and the fact that $D_{\n\m}^{</>}$ is symmetric in $\m\n$, we easily see the two diagrams give identical contributions.
One may wonder about the contribution when both vertices are resummed. We show by enumeration in Appendix E that such contribution is not allowed.

Note that $p_0=p$ with our choice. It is useful to rearrange our results above. Using the reciprocal relations \eqref{reciprocal_f} and \eqref{reciprocal_p}, we can deduce diagrammatically analogous relation holds for fermion self-energy from the pinching kinematical region.
\begin{align}
\S^>(P)=\overline{\S}^<(-P)={\S}^<(-P).
\end{align}
In the second equality, we have removed the overline because we have dropped the subleading mass next to $\slashed{P}$ and mass always appears in squares elsewhere. We can then rewrite
\begin{align}
&\tr[\S^>(P)S^<(P)]\vert_{\text{2nd}}=\tr[\S^<(-P)S^>(-P)]\vert_{\text{1st}},\no
&\tr[\S^<(P)S^>(P)]\vert_{\text{2nd}}=\tr[\S^>(-P)S^<(-P)]\vert_{\text{1st}},
\end{align}
where the subscripts $\text{1st}$ and $\text{2nd}$ denote contributions from the first and second diagrams in Fig.\ref{se-pinch2} respectively.
This allows us to reexpress the contributions in the region $p_0=p$ from two diagrams to the counterpart in the region $p_0=\pm p$ from a single diagram:
\begin{align}\label{SS_sum}
&\tr[\S^>(P)S^<(P)-\S^<(P)S^>(P)]_{\text{1st}+\text{2nd}}\no
=&\tr[\S^>(P)S^<(P)-\S^<(P)S^>(P)-\S^>(-P)S^<(-P)+\S^<(-P)S^>(-P)]_{\text{1st}}
\end{align}
In fact, the RHS of \eqref{SS_sum} is closely related to \eqref{coll_p}. Note that $\P^{MN</>}(K)=\int_P G^{MN</>}(P+K,P)$. We would have the collision term of photon if the momentum integration is done over $P$ instead of $K$. To establish the relation, we first integrate $G^{MN</>}$ over transverse components of momentum so that the resulting collision term has natural interpretation as collinear $1\leftrightarrow2$ processes. This can be conveniently done by converting $d^2k_\pp$ to $d^2p_\pp$. The key observation is that the geometrically invariant quantity here is the opening angle $\th$ between $\vp$ and $\vk$, which is bounded by the integrand as $\th\lesssim\frac{m_\g}{p}\sim O(e)$. Using spherical coordinates, we can easily show
\begin{align}
\int\frac{d^2k_\pp}{(2\p)^2}=\frac{k^2}{p^2}\int\frac{d^2p_\pp}{(2\p)^2}.
\end{align}

The remaining complication is in the integration of $k_0$.
On one hand, $G^{MN</>}$ contains $\d(p_0-p_\pr\e(k_0))$ according to the pinching mechanism. On the other hand $\d(K^2)$ from $D_{MN}^{</>}=\frac{1}{2k}\(\d(k_0-k)+\d(k_0+k)\)$ receives contribution from two poles. For $k_0=k$, it gives the following contribution
\begin{align}\label{coll_f1}
&\int_KG^{MN>}D_{MN}^<=\int_0^\infty\frac{dk}{2\p}\int\frac{d^2p_\pp}{(2\p)^2}\frac{k}{2p^2}f_e(p_\pr)(1-f_e(p_\pr+k))2\text{Re}[\c(P)]2\p\d(p_0-p_\pr)\no
&\frac{\((p_\pr+k)^2+p_\pr^2\)p_\pp^2}{\(p_\pr(p_\pr+k)\)^2}2\p(1+f_\g(K)).
\end{align}
For $k_0<-k$, we have instead 
\begin{align}\label{coll_f2}
&\int_KG^{MN>}D_{MN}^<=\int_0^\infty\frac{dk}{2\p}\int\frac{d^2p_\pp}{(2\p)^2}\frac{k}{2p^2}f_e(-p_\pr)(1-f_e(-p_\pr-k))2\text{Re}[\c(P)]2\p\d(p_0+p_\pr)\no
&\frac{\((-p_\pr-k)^2+p_\pr^2\)p_\pp^2}{\(-p_\pr(-p_\pr-k)\)^2}2\p(1+f_\g(K)).
\end{align}
Note that we need to have $p_\pr>0(p_\pr<0)$ for $k_0>0(k_0<0)$ respectively in order to give the pole contribution at $p_0=p$ for fermion. We can then combine \eqref{coll_f1} and \eqref{coll_f2} to write
\begin{align}\label{coll_f}
&\tr[\S^>(P)S^<(P)-\S^<(P)S^>(P)]=\int\frac{dk_\pr}{2\p}\int\frac{d^2p_\pp}{(2\p)^2}\frac{k}{2p^2}f_e(p)(1-f_e(p+k_\pr))\no
&\times2\text{Re}[\c(P)]2\p\d(p_0-p)\frac{\((p+k_\pr)^2+p^2\)p_\pp^2}{\(p(p+k_\pr)\)^2}2\p(1+f_\g(K)).
\end{align}
\eqref{coll_f} is in agreement with Eq.(5.3) of \cite{Arnold:2002zm} when specialized to $U(1)$ gauge group.

\section{Quantum correction}\label{sec_hbar}

Now we proceed to the quantum correction, which captures dynamics of spin polarization. We first consider the quantum correction to fermion Wigner function $S^{<(1)}$, which enters two equations from expansion in $\hb$ at different orders
\begin{align}\label{S1_eqn}
&\frac{i}{2}\slashed{\pd}S^{<(0)}+\frac{\slashed{P}-m}{\hb}S^{<(1)}=\frac{i}{2}\(\S^{>(0)}S^{<(0)}-\S^{<(0)}S^{>(0)}\),\no
&\frac{i}{2}\slashed{\pd}S^{<(1)}+\frac{\slashed{P}-m}{\hb}S^{<(2)}=\frac{i}{2}\(\S^{>(1)}S^{<(0)}-\S^{<(1)}S^{>(0)}+\S^{>(0)}S^{<(1)}-\S^{<(0)}S^{>(1)}\)\no
&-\frac{\hb}{4}\(\{\S^{>(0)},S^{<(0)}\}_\pb-\{\S^{<(0)},S^{>(0)}\}_\pb\).
\end{align}
Recall that the trace of the first equation has been used to derive the classical kinetic equation. The off-diagonal components serve as a constraint equation for $S^{<(1)}$. It determines the non-dynamical part of $S^{<(1)}$. The second equation is dynamical for $S^{<(1)}$. Its role is similar to that of the first equation to $S^{<(0)}$. We expect that the second equation would give rise to an additional axial component of $S^{<(1)}$ in the form $a^\m f_A$ with $f_A\sim O(\frac{\hb\pd_Xf}{\L})$. Note that although $f_A$ is excluded by the parity invariant assumption at lowest order in $\hb$, it is inevitable at next order for restoring frame independence of the Wigner function \cite{Hidaka:2016yjf,Hidaka:2017auj}. This is a massive generalization of the side-jump effect in the chiral limit \cite{Chen:2014cla,Chen:2015gta}, see also \cite{Yang:2020hri,Liu:2021uhn}. We note that while it is possible to determine the dynamical part of axial component of Wigner function in global equilibrium by principle of frame independence \cite{Gao:2018jsi}, the generalization to local equilbrium case can be non-trivial. In this paper, we focus on the non-dynamical part of the axial component and leave the dynamical part for separate studies.


By solving the constraint equation in \eqref{S1_eqn}, we obtain the non-dynamical part of both axial and tensor components:
\begin{align}\label{S1}
S^{<(1)}(P)=\g^5\g_\m{\cal A}^\m+\frac{i[\g_\m,\g_\n]}{4}{\cal S}^{\m\n},
\end{align}
where
\begin{align}\label{AS}
&{\cal A}^\m=-2\p\hb\e(P\cdot u)\frac{\e^{\m\n\r\s}P_\r u_\s{\cal D}_\n f_e}{2(P\cdot u+m)}\d(P^2-m^2),\no
&{\cal S}^{\m\n}=-2\p\hb\e(P\cdot u)\frac{{\cal D}_{[\m} P_{\n]}f_e-m u_{[\m} {\cal D}_{\n]}f_e-P_{[\m} u_{\n]}{\cal D}_m}{2(P\cdot u+m)}\d(P^2-m^2).
\end{align}
The collisional effect is taken into account in the definitions ${\cal D}_\n=\pd_\n-\S_\n^>-\S_\n^<\frac{1-f_e}{f_e}$ and ${\cal D}_m=\S_m^>+\S_m^<\frac{1-f_e}{f_e}$ with $\S_\n^{>/<}=\frac{1}{4}\tr[\S^{>/<}\g_\n]$ and $\S_m^{>/<}=\frac{1}{4}\tr[\S^{>/<}]$. Since ${\cal A}^\m$ gives rise to spin polarization upon frequency integration, see for example \cite{Becattini:2013fla}, the collisional contribution to spin polarization is naturally can be readily studied for any given solution to classical kinetic theory.

Let us elaborate on the nature of $\hb$ expansion. Recall that in early studies of transport coefficients, $\hb$ is set to unity and gradient expansion is used in solving the classical kinetic equation. The gradient expansion of the distribution $f$ read: $f=f_{(0)}+f_{(1)}+\cdots$. Restoring $\hb$ by dimension, we have $f_{(0)}\sim O(1)$, $f_{(1)}\sim\frac{\hb\pd_Xf_{(0)}}{\L}$. It is clear that the gradient always comes with $\hb$, therefore partial corrections at higher order in $\hb$ is already present in solution to classical kinetic theory. This type of $\hb$ correction distinguishes from the one discussed above in that the former does not contribute to spin polarization while the latter does. We shall still use the terminology classical and quantum as a separation between spin unpolarized and polarized sectors of kinetic theory.


We move on to the quantum correction to photon Wigner function $D_{\n\r}^{<(1)}$ with the same logic in mind. Keeping only the non-dynamical equation, we have
\begin{align}\label{D1_eqn}
&\(-P^2g^{\m\n}+P^\m P^\n-\frac{1}{\x}P^{\m\a}P^{\n\b}P_\a P_\b\)D_{\n\r}^{<(1)}+\frac{i}{2}\big(-2P\cdot\pd g^{\m\n}+\pd^\m P^\n+\pd^\n P^\m \no
&-\frac{1}{\x}P^{\m\a}P^{\n\b}(\pd_\a P_\b+\pd_\b P_\a)\big)D_{\n\r}^{<(0)}
=\frac{i\hb}{2}\(\P^{\m\n>(0)}D_{\n\r}^{<(0)}-\P^{\m\n<(0)}D_{\n\r}^{>(0)}\),
\end{align}
With the help of the gauge fixing term, we can solve \eqref{D1_eqn} by the following inversion
\begin{align}\label{inversion}
\(\frac{u_\m u_\l}{{\bf p}^2}+\frac{P_{\m\l}^T}{P^2}-\x\frac{P_\m P_\l}{{\bf p}^4}\)\(-P^2g^{\m\n}+P^\m P^\n-\frac{1}{\x}P^{\m\a}P^{\n\b}P_\a P_\b\)=\d_\l^\n,
\end{align}
with ${\bf p}^2=-P^2+(P\cdot u)^2$.
Note that $\(\frac{u_\m u_\l}{{\bf p}^2}+\frac{P_{\m\l}^T}{P^2}-\x\frac{P_\m P_\l}{{\bf p}^4}\)$ gives the Coulomb gauge propagator at $\x=0$. Multiplying it to \eqref{D1_eqn}, we obtain
\begin{align}\label{D1_sol}
&D_{\l\r}^{<(1)}=-\(\frac{u_\m u_\l}{{\bf p}^2}+\frac{P_{\m\l}^T}{P^2}-\x\frac{P_\m P_\l}{{\bf p}^4}\)\frac{i}{2}\big(-2P\cdot\pd g^{\m\n}+\pd^\m P^\n+\pd^\n P^\m-\frac{1}{\x}P^{\m\a}P^{\n\b}\no
&\times(\pd_\a P_\b+\pd_\b P_\a)\big)D_{\n\r}^{<(0)}
+\(\frac{u_\m u_\l}{{\bf p}^2}+\frac{P_{\m\l}^T}{P^2}-\x\frac{P_\m P_\l}{{\bf p}^4}\)\frac{i\hb}{2}\(\P^{\m\n>(0)}D_{\n\r}^{<(0)}-\P^{\m\n<(0)}D_{\n\r}^{>(0)}\).
\end{align}
We first show terms $\propto P_{\m\l}^T$ vanish by classical kinetic equations. To see that, we note $D_{\n\r}^{</>(0)}\propto P_{\n\r}^T$. The only remaining terms on the RHS are
\begin{align}
\frac{P_{\m\l}^T}{P^2}\frac{i}{2}\(2P\cdot\pd g^{\m\n}D_{\n\r}^{<(0)}+\hb\P^{\m\n>(0)}D_{\n\r}^{<(0)}-\hb\P^{\m\n<(0)}D_{\n\r}^{>(0)}\)\propto P_{\l\r}^T.
\end{align}
Since the tensor structure is unique, we can extract its coefficient function by contracting with $P^{\l\r}_T$, which vanishes by classical kinetic equation. The other terms give the following result in the limit $\x\to0$
\begin{align}\label{D1_half}
&D_{\l\r}^{<(1)}=-\frac{iu_\l(P\cdot u)\pd^\n D_{\n\r}^{<(0)}}{2{\bf p}^2}-\frac{iP_\l P^{\n\b}\pd_\b D_{\n\r}^{<(0)}}{2{\bf p}^2}+\frac{i\hb u_\l u_\m\(\P^{\m\n>(0)}D_{\n\r}^{<(0)}-\P^{\m\n<(0)}D_{\n\r}^{>(0)}\)}{2{\bf p}^2}\no
&=-\frac{iP_{\l\a}P^{\n\b}P^\a\pd_\b D_{\n\r}^{<(0)}}{2{\bf p}^2}+\frac{i\hb u_\l u_\m\(\P^{\m\n>(0)}D_{\n\r}^{<(0)}-\P^{\m\n<(0)}D_{\n\r}^{>(0)}\)}{2{\bf p}^2}.
\end{align}
In the second line, we have used $\pd^\n D_{\n\r}^{<(0)}=-P^{\n\b}\pd_\b D_{\n\r}^{<(0)}$. However, \eqref{D1_half} cannot be the correct quantum correction to Wigner function. Since $D_{\n\r}^{<}$ is hermitian, a purely imaginary $D_{\n\r}^{<(1)}$ in \eqref{D1_half} indicates it should be anti-symmetric in indices \cite{Hattori:2020gqh}, which is obviously not satisfied by \eqref{D1_half}. The resolution is simple: since we consider on-shell photon, the operator $-P^2g^{\m\n}+P^\m P^\n-\frac{1}{\x}P^{\m\a}P^{\n\b}P_\a P_\b$ contains zero mode. \eqref{D1_half} can be modified by zero mode contribution. We can make \eqref{D1_half} hermitian by adding its own hermitian conjugate
\begin{align}\label{D1_half2}
D_{\l\r}^{<(1)}=\frac{iP_{\r\a}P^{\n\b}P^\a\pd_\b D_{\n\l}^{<(0)}}{2{\bf p}^2}+\frac{i\hb u_\r u_\m\(\P^{\m\n>(0)}D_{\n\l}^{<(0)}-\P^{\m\n<(0)}D_{\n\l}^{>(0)}\)}{2{\bf p}^2}.
\end{align}
It turns out \eqref{D1_half2} is at the same time a zero mode of $-P^2g^{\m\n}+P^\m P^\n-\frac{1}{\x}P^{\m\a}P^{\n\b}P_\a P_\b$ using $D_{\n\l}^{<(0)}\propto P_{\n\l}^T\d(P^2)$. Adding up \eqref{D1_half} and \eqref{D1_half2}, we find the quantum correction\footnote{The appearance of a factor of $\hb$ is consistent with dimension of photon self-energy $\P^{\m\n}(P)\sim \text{length}^{-2}$.}
\begin{align}\label{D1_final}
&D_{\l\r}^{<(1)}=-\frac{iP_{\l\a}P^{\n\b}P^\a\pd_\b D_{\n\r}^{<(0)}}{2(-P^2+(P\cdot u)^2)}+\frac{i\hb u_\l u_\m\(\P^{\m\n>(0)}D_{\n\r}^{<(0)}-\P^{\m\n<(0)}D_{\n\r}^{>(0)}\)}{2(-P^2+(P\cdot u)^2)}-(\l\leftrightarrow\r)\no
&=-2\p\e(P\cdot u)\d(P^2)\frac{iP_{\l\a}P^{\n\b}P^\a\pd_\b P_{\n\r}^T f_\g(P)}{2(-P^2+(P\cdot u)^2)}+2\p\e(P\cdot u)\d(P^2)P_{\n\r}^T\times\no
&\frac{i\hb u_\l u_\m\(\P^{\m\n>(0)}f_\g(P)-\P^{\m\n<(0)}(1+f_\g(P))\)}{2(-P^2+(P\cdot u)^2)}-(\l\leftrightarrow\r).
\end{align}
In the collisionless limit, \eqref{D1_final} agrees with \cite{Hattori:2020gqh}. To see the equivalence, we note that counterpart in \cite{Hattori:2020gqh} can be obtained by replacing $P_{\n\r}^T$ in the first term by $P_{\n\r}$. The difference from the replacement is proportional to
\begin{align}
P_{\l\a}P^\a P_\m P^{\m\b}\pd_\b P_{\r\s}P^\s-(\l\leftrightarrow\r),
\end{align}
which vanishes identically by the anti-symmetrization in indices. \eqref{D1_final} also contains collisional contribution, which has the form of $\P^{\m\n}$ sandwiched between $u_\l u_\m$ and $P^T_{\n\r}$. In an isotropic medium, $\P^{\m\n}=P_L^{\m\n}\P_L+P_T^{\m\n}\P_T$, so this contribution vanishes identically by orthogonality condition of projectors. In a generic off-equilbrium medium, for example one with shear gradient, more structures are allowed in the self-energy, so the collisional contribution is in general nonvanishing. Unlike in the case of screening, the effect of gradient on the quantum correction here is leading order.


\section{Outlook}\label{sec_outlook}

We have derived a quantum kinetic theory for QED by assuming parity invariance. At lowest order in $\hb$, it generalizes the well-known classical kinetic theory to massive case. We have also found a non-dynamical quantum correction to the Wigner function at $O(\hb)$, which gives the spin polarization of fermions and photons. Several interesting extensions of this work can be studied:

First of all, relaxing the parity invariant constraint would introduce more degrees of freedom to the kinetic theory. It makes the distribution functions spin-dependent, allowing us to study the the spin evolution within kinetic theory.
These are of particular interest in the physics like chiral magnetic effect, in which local parity violation is also present.

Secondly, the notion of locality can be further explored in the kinetic theory. We have different scales for collisions with $1/e^2\L$ corresponds to the effective range of inelastic collisions and $1/e\L$ corresponds to the range of elastic collisions. In our case, both are local because of our larger coarse-graining scale $1/e^4\L$. It would be interesting to look at other possibilities, for example, choosing $1/e\L$ as coarse-graining scale would lead to non-local inelastic collisions, allowing us to study transfer between orbital and spin angular momenta. With this finer notion of locality, we also need to take into account the electromagnetic field from fluctuations.

Finally it is clearly desirable to extend the present study to QCD case and determine the corresponding spin transport coefficients. This would shed light on the applicability of spin hydrodynamics in the system of spinning quark-gluon plasma produced heavy ion collisions.

\section{Acknowledgments}
We are grateful to Wei-jie Fu, Jian-hua Gao, Defu Hou, Shi Pu, Xin-li Sheng and Ziyue Wang for useful discussions. We are also grateful to anonymous referee for insightful criticisms on earlier versions of the paper. This work is in part supported by NSFC under Grant Nos 12075328, 11735007 and 11675274.

\appendix

\section{From self-energies to collision term}

In this appendix, we show how \eqref{SS}, \eqref{SS2}, \eqref{PD}, \eqref{PD_v}, \eqref{SS_v1} and \eqref{SS_v2} can be reduced to collision term in Boltzmann equation for QED.

We start with \eqref{SS} and focus on the case $p_0>0$, i.e. fermion in the initial state. The case $p_0<0$ can be deduced by crossing symmetry. There are in total $8$ ways in choosing the sign of $k_0$, $k_0'$ and $p_0'$. Only the following three are kinematically allowed: $k_0>0,\,k_0'>0,\, p_0'>0$; $k_0<0,\,k_0'<0,\, p_0'>0$ and $k_0<0,\,k_0'>0,\, p_0'<0$. The second and third cases are related to the first one by crossing symmetry. Let us consider the first case. The kinematical region implies it corresponds to square of s-channel Compton. To see that, we use
\begin{align}\label{pol_sum}
P_{\n\m}^T(K)=\sum_i\e_\n^i(K)\e_\m^{i*}(K),
\end{align}
and similarly for $P_{\s\r}^T(K')$, which relate the projector to sum over initial/final states of photon. We can further use the spin sum formula for $\slashed{P}+m$ from $S^<(P)$
\begin{align}\label{spin_sum}
\slashed{P}+m=\sum_s u_s(P)\bar{u}_s(P),
\end{align}
and similarly for $\slashed{P'}+m$ from $S^>(P')$. Recall in Section.\ref{sec_KB} that trace will be taken in deriving dynamical equation for $f_e$. With all the rewritings above and cyclic property of trace, we have from $\tr[\eqref{SS}]$\footnote{An overall numerical factor $(2\p)^4$ is suppressed.}
\begin{align}
\sum_{s,t,i,j}&\tr[\bar{u}_s(P)\g^\m S_{11}(K+P)\g^\r u_t(P')\e^i_{\s}(K')\e^{j}_{\m}(K)\bar{u}_t(P')\g^\s S_{22}(K+P)\g^\n u_s(P)\e^{i*}_{\r}(K')\e^{j*}_{\n}(K)]\no
\times &f_e(P)f_\g(K)(1-f_e(P'))(1+f_\g(K'))\d(P^2)\d(P'{}^2)\d(K^2)\d(K'{}^2),
\end{align}
which is clearly identified as square of s-channel Compton scattering. By crossing symmetry, we obtain square of t-channel Compton scattering and t-channel annihilation for the other two cases. The action of crossing symmetry on distribution functions use the constraints $f_e(P)+f_e(-P)=1$ and $f_\g(P)+f_\g(-P)=1$ to convert distribution functions of particles to that of anti-particles.

\eqref{SS2} can be rewritten similarly. For the case $p_0>0$, $k_0>0$, $k_0'>0$ and $p_0'>0$, we also use \eqref{pol_sum}, \eqref{spin_sum} to rewrite $\tr[\eqref{SS2}]$ as
\begin{align}
&\sum_{s,t,i,j}\tr[\bar{u}_s(K')\g^\b u_t(P')\bar{u}_t(P')\g^\a u_s(K')]\tr[\bar{u}_i(P)\g^\m u_j(K)\bar{u}_j(K)\g^\n u_i(P)]D_{\n\a}^{11}(K-P)\times\no
& D_{\b\m}^{22}(K-P)f_e(P)f_e(P')(1-f_e(K))(1-f_e(K'))\d(P^2)\d(P'{}^2)\d(K^2-m^2)\d(K'{}^2-m^2),
\end{align}
which corresponds to square of t-channel Coulomb scattering between fermions. By crossing symmetry, we can obtain also the squares of s/t-channel Coulomb scattering between fermion and anti-fermion.

Now we turn to \eqref{PD}. In fact, $\tr[\eqref{SS}]$ and \eqref{PD} are simply related by relabeling of momenta $P\leftrightarrow K$. It follows that \eqref{PD} also correspond to squares of s/t-channel Compton scattering and t-channel annihilation.

The identification with interference terms proceeds similarly. For vertex correction to photon self-energy \eqref{PD_v}, we can use cyclic property of trace to rewrite \eqref{PD_v} up to overall factors of $f_\g(P)f_e(P')(1-f_e(K))(1+f_\g(K'))\d(P^2)\d(P'{}^2)\d(K^2-m^2)\d(K'{}^2-m^2)$ as
\begin{align}
&\sum_{s,t,i,j}\tr[\bar{u}_s(K)\g^\m S_{22}(P-K)\g^\r u_t(P')]\e_\r^i(Q)\e_\m^{j*}(P)\no
&\times\tr[\bar{u}_t(P')\g^\n S_{11}(P+P')\g^\s u_s(K)]\e_\s^{i*}(Q)\e_\n^{j}(P)\no
&=\sum_{s,t,i,j}\(\tr[\bar{u}_t(P')\g^\r S_{11}(P-K)\g^\m u_s(K)]\e_\r^{i*}(Q)\e_\m^{j}(P)\)^*\no
&\times\tr[\bar{u}_t(P')\g^\n S_{11}(P+P')\g^\s u_s(K)]\e_\s^{i*}(Q)\e_\n^{j}(P).
\end{align}
This is the product of the s-channel amplitude with the complex conjugate of t-channel amplitude for Compton scattering, to be abbreviated $st^*$ of Compton.
By crossing symmetry, we can easily obtain the $s^*t$ of Compton and $t^*u$ of annihilation. Note that we have only half of the interference terms for annihilation.

For the vertex correction to fermion self-energy, we look at \eqref{SS_v1} first. Using \eqref{reciprocal_f}, we have
\begin{align}
S^>(-P')=\overline{S}^<(P').
\end{align}
Note that $\overline{S}$ has sign of mass flipped. We can represent the corresponding Dirac structure by the spin sum of anti-fermions
\begin{align}\label{spin_sum2}
\slashed{P'}-m=\sum_sv_s(P')\bar{v}_s(P').
\end{align}
We can then rewrite the trace of \eqref{SS_v1} up to $f_e(P)f_{{e}}(P')(1+f_\g(K))(1+f_\g(K'))\d(P^2-m^2)\d(P'{}^2-m^2)\d(K^2)\d(K'{}^2)$ as
\begin{align}
&\sum_{s,t,i,j}\tr[\bar{u}_s(P)\g^\m S_{22}(P-K)\g^\n v_t(P')]\e_\m(K)\e_\n^*(K')\no
\times&\tr[\bar{v}_t(P')\g^\r S_{11}(P-K')\g^\s u_s(P)]\e_\r^*(K)\e_\s(K')\no
=&\sum_{s,t,i,j}\(\tr[\bar{v}_t(P')\g^\n S_{11}(P-K)\g^\m u_s(P)]\e_\m^*(K)\e_\n(K')\)^*\no
\times&\tr[\bar{v}_t(P')\g^\r S_{11}(P-K')\g^\s u_s(P)]\e_\r^*(K)\e_\s(K'),
\end{align}
which corresponds to half of the interference term of annihilation. By crossing symmetry, we can obtain interference terms of Compton scattering.

Finally we turn to \eqref{SS_v2}, which has slightly different momenta labeling from the other cases, with $P$, $P'$, $P-K$ and $P-K'$ on-shell instead. We consider the kinematical region with $p_0'<0,\,p_0-k_0>0,\,p_0-k_0'>0$. Using similar procedure as before, we rewrite \eqref{SS_v2} up to overall factors of $f_e(P)f_{{e}}(P')(1+f_\g(P-K))(1+f_\g(P-K'))\d(P^2-m^2)\d(P'{}^2-m^2)\d((P-K)^2-m^2)\d((P-K')^2-m^2)$ as
\begin{align}
&\sum_{s,t,i,j}\tr[\bar{u}_s(P)\g^\m u_t(P-K)]\tr[\bar{u}_i(-P')\g^\r u_j(P-K')]D_{\m\r}^{22}(K)\no
\times&\tr[\bar{u}_t(P-K)\g^\n u_i(-P')]\tr[\bar{u}_j(P-K')\g^\s u_s(P)]D_{\n\s}^{11}(K')\no
=&\sum_{s,t,i,j}\(\tr[\bar{u}_t(P-K)\g^\m u_s(P)]\tr[\bar{u}_j(P-K')\g^\r u_i(-P')]D_{\m\r}^{11}(K)\)^*\no
\times&\tr[\bar{u}_t(P-K)\g^\n u_i(-P')]\tr[\bar{u}_j(P-K')\g^\s u_s(P)]D_{\n\s}^{11}(K').
\end{align}
This gives the product of t-channel amplitude with complex conjugate of u-channel amplitude of Coulomb scattering between fermions ($tu^*$). By crossing symmetry, we can also obtain $st^*$ and $s^*t$ of Coulomb scattering between fermion and anti-fermion.

\section{Evaluation of self-energies}

In this appendix, we evaluate the retarded self-energy of fermions and photons, which are needed for determination of thermal masses in the main text. The retarded self-energies have the following representation in $ra$ basis: $S_R=iS_{ra}$ and $D^R_{\m\n}=iD^{ra}_{\m\n}$, which allows us to perform the calculation in $ra$ basis. 

\begin{figure}
\includegraphics[width=0.45\textwidth]{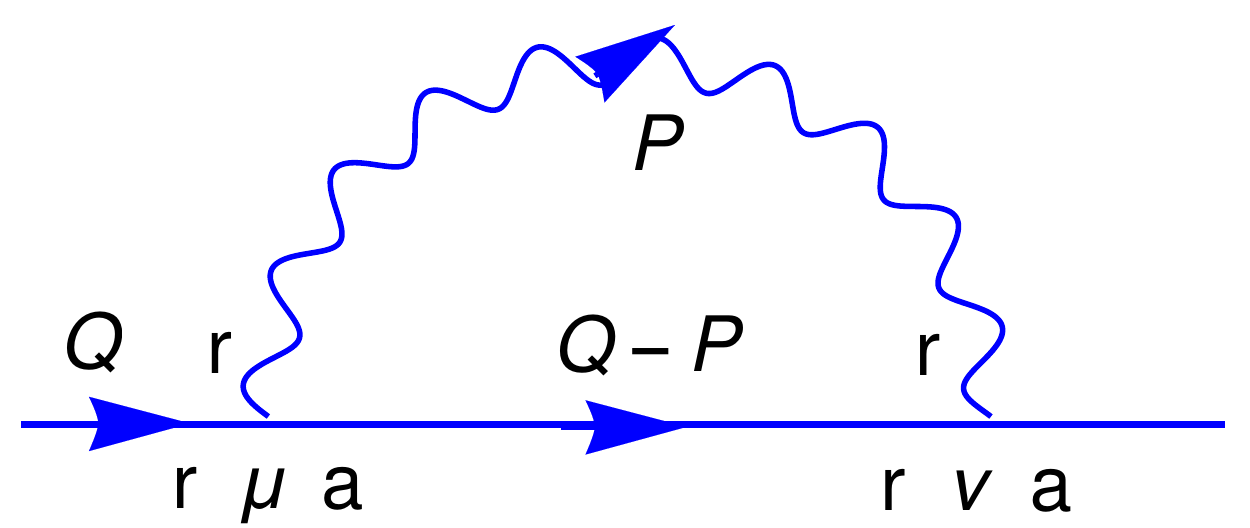}
\includegraphics[width=0.45\textwidth]{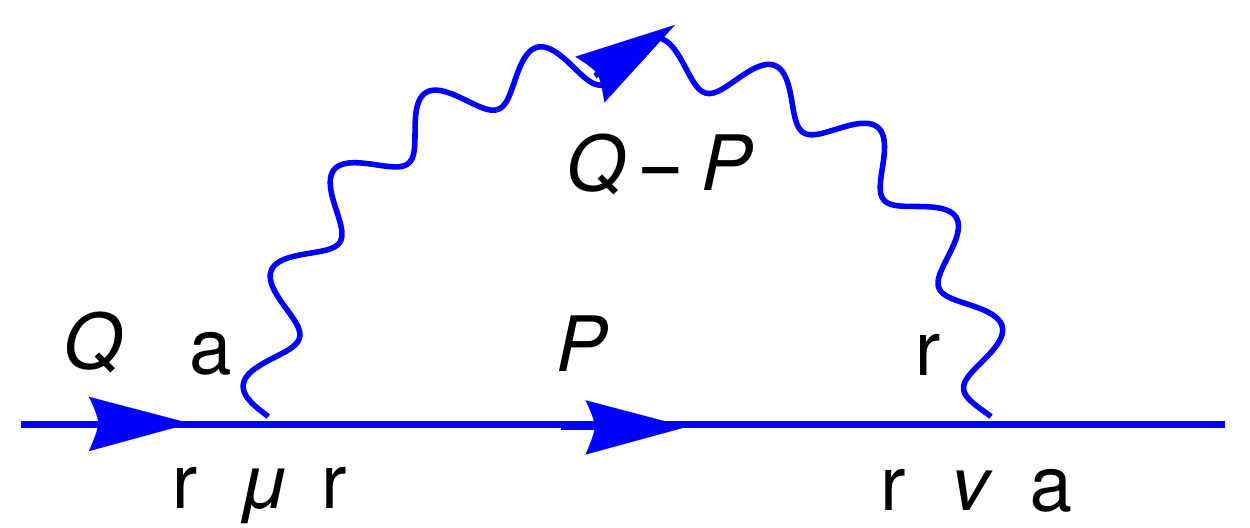}
\caption{\label{Rse-fermion}One-loop fermion self-energy $\S_{ra}$. Vertices can have labelings $rra$ and $aaa$.}
\end{figure}

We first evaluate the fermion self-energy, for which we need in two different kinematical regions: soft off-shell momenta (for screening effect in elastic collisions) and hard on-shell momenta (for inelastic collisions). In both cases, we also require $m\lesssim e\L$ in order for screening or inelastic collisions to be relevant. This condition allows us to drop $m$ in the evaluation of fermion self-energy. It is known from explicit calculations that the fermion self-energies in equilibrium are the same in Coulomb gauge \cite{Blaizot:2001nr} and Feynman gauge \cite{Bellac:2011kqa}. The agreement is expected to hold for off-equilibrium case at the lowest order because the structures of propagators do not change from equilibrium to off-equilibrium case. Below we will proceed in Feynman gauge. Fig.~\ref{Rse-fermion} shows the self-energy diagrams in $ra$ basis, which give
\begin{align}
\S_{ra}(Q)=-e^2\int_P\g^\m S_{ar}(Q-P)\g^\n D^{rr}_{\n\m}(P)-e^2\int_P\g^\m S_{rr}(P)\g^\n D_{\n\m}^{ar}(Q-P).
\end{align}
The relevant propagators in $ra$ basis are give by
\begin{align}
S_{ar}(P)&=\frac{i(\slashed{P}+m)}{(p_0-i\e)^2-p^2-m^2},\no
S_{rr}(P)&=\(\frac{1}{2}-f_e(P)\)(\slashed{P}+m)2\p\e(p_0)\d(P^2-m^2),\no
D_{\m\n}^{ar}(P)&=\frac{-ig_{\m\n}}{(p_0-i\e)^2-p^2},\no
D_{\m\n}^{rr}(P)&=-g_{\m\n}\(\frac{1}{2}+f_\g(P)\)2\p\e(p_0)\d(P^2).\nonumber
\end{align}
As we argued above, we can drop $m$ in $S_{ar}$ and $S_{rr}$. Furthermore, we can drop $Q^2$ in the denominators of $S_{ar}(Q-P)$ and $D_{\n\m}^{ar}(Q-P)$, because either $Q$ is soft off-shell $Q^2\ll Q\cdot P$ or hard on-shell $Q^2=0$. We thus have
\begin{align}\label{Sigma_ra}
i\S_{ra}(Q)&=-e^2\int_P\frac{-2(\slashed{Q}-\slashed{P})}{-2P\cdot Q}\(\frac{1}{2}+f_\g(P)\)2\p\e(p_0)\d(P^2)\no
&-e^2\int_P\frac{-2\slashed{P}}{-2P\cdot Q}\(\frac{1}{2}-f_e(P)\)2\p\e(p_0)\d(P^2),
\end{align}
where we have dropped $i\e$ in the denominators because $Q-P$ is off-shell. Separating the cases of particles and anti-particles, we can expand \eqref{Sigma_ra} as
\begin{align}
i\S_{ra}(Q)&=-e^2\int\frac{d^3p}{(2\p)^3}\frac{1}{2p}\(\frac{-2(\slashed{Q}-\slashed{P})}{-2P\cdot Q}\(\frac{1}{2}+f_\g(\vp)\)-\frac{-2(\slashed{Q}-\slashed{\bar{P}})}{-2\bar{P}\cdot Q}\(\frac{1}{2}-f_\g(-\vp)\)\)\no
&-e^2\int\frac{d^3p}{(2\p)^3}\frac{1}{2p}\(\frac{-2\slashed{P}}{-2P\cdot Q}\(\frac{1}{2}-f_e(\vp)\)-\frac{-2\slashed{\bar{P}}}{-2\bar{P}\cdot Q}\(-\frac{1}{2}-f_{\bar{e}}(-\vp)\)\),
\end{align}
where $\bar{P}=(-p_0,\vp)$ and $f_{\bar{e}}$ is the distribution function of anti-fermions defined below \eqref{KE_anti}.
We can then use a change a variable $\vp\to-\vp$ to arrive at
\begin{align}
i\S_{ra}(Q)&=-e^2\int\frac{d^3p}{(2\p)^3}\frac{1}{2p}\(\frac{-2(\slashed{Q}-\slashed{P})}{-2P\cdot Q}\(\frac{1}{2}+f_\g(\vp)\)-\frac{-2(\slashed{Q}+\slashed{P})}{2P\cdot Q}\(-\frac{1}{2}-f_\g(\vp)\)\)\no
&-e^2\int\frac{d^3p}{(2\p)^3}\frac{1}{2p}\(\frac{-2\slashed{P}}{-2P\cdot Q}\(\frac{1}{2}-f_e(\vp)\)-\frac{2\slashed{P}}{2P\cdot Q}\(\frac{1}{2}+f_{\bar{e}}(\vp)\)\).
\end{align}
Dropping the $\frac{1}{2}$ in the brackets, which correspond to the vacuum contributions, we end up with
\begin{align}\label{Sigma_ra_final}
i\S_{ra}(Q)=e^2\int\frac{d^3p}{(2\p)^3}\frac{1}{2p}\frac{\slashed{P}}{P\cdot Q}\(2f_\g(\vp)+f_e(\vp)+f_{\bar{e}}(\vp)\).
\end{align}
Since we require $m\lesssim e\L$ and $Q$ can be either hard on-shell or soft off-shell, including mass in the evaluation of fermion self-energy would only lead to a correction at order $O(e^3\L^2/Q)$.

\begin{figure}
\includegraphics[width=0.45\textwidth]{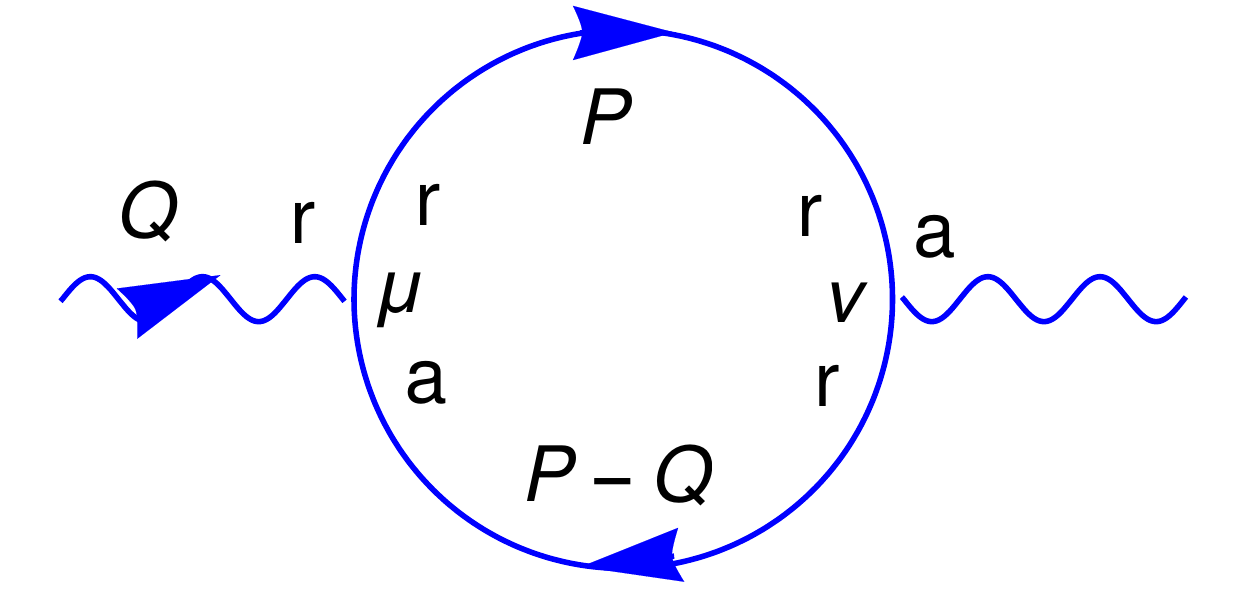}
\includegraphics[width=0.45\textwidth]{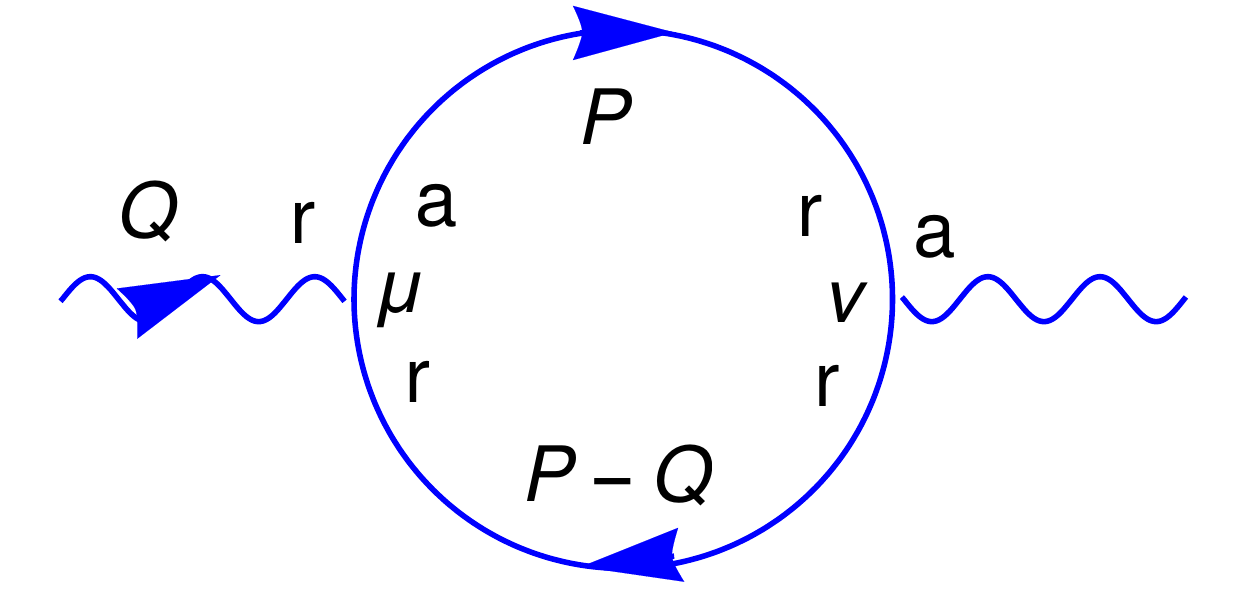}
\caption{\label{Rse-photon}One-loop hard photon self-energy $\P_{ra}$. Vertices can have labelings $rra$ and $aaa$.}
\end{figure}

Next we turn to the retarded photon self-energy, for which we also need in two different kinematical regions: soft off-shell momenta (for screening in elastic collisions) and hard on-shell momenta (for inelastic collisions). However, there is one small difference from the fermion self-energy. While in case of inelastic collisions, photon self-energy is never needed for very massive fermion $m\gg e\L$ because this would make inelastic scattering itself irrelevent, it is always needed for screening in elastic collisions for arbitrary fermion mass. So we will not drop $m$ as in the following evaluation of photon self-energy. Since the photon self-energy at leading order in coupling is gauge invariant. We can also evaluate it in Feynman gauge.
The corresponding diagrams are shown in Fig.~\ref{Rse-photon}, which give
\begin{align}
\P_{\m\n}^{ra}&=e^2\int_P\tr[\g^\m S_{rr}(P)\g^\n S_{ra}(P-Q)]+e^2\int_P\tr[\g^\m S_{ar}(P)\g^\n S_{rr}(Q-P)]\no
&=e^2\int_P\tr[\g^\m S_{rr}(P)\g^\n S_{ra}(P-Q)]+(Q\to-Q,\m\leftrightarrow\n).
\end{align}
The contribution from the second diagram is related the counterpart from the first one by the replacement $Q\to-Q,\m\leftrightarrow\n$. This can be shown by using the property $S_{ar}(P)=S_{ra}(P)$ valid for off-shell momentum and relabeling of momentum $P\to Q-P$.
Keeping fermion mass in the propagators for the photonic case, we evaluate the trace as
\begin{align}\label{trace}
\tr[\g^\m(\slashed{P}+m)\g^\n(\slashed{P}-\slashed{Q}+m)]=8P^\m P^\n-4(P^\m Q^\n+P^\n Q^\m)+4g^{\m\n}P\cdot Q.
\end{align}
Note that the $m$-dependence drops by the on-shell condition of $P$. It turns out that the leading contribution from the term $P^\m P^\n$ vanishes upon combination of the two diagram, thus we cannot simply drop the $Q^2$ in $S_{ra}(P-Q)$ as in the fermionic case. Instead we should approximate $S_{ra}(P-Q)$ as
\begin{align}\label{Q2_exp}
S_{ra}(P-Q)\simeq\frac{i(\slashed{P}-\slashed{Q}+m)}{-2P\cdot Q}\(1+\frac{Q^2}{2P\cdot Q}\).
\end{align}
The remaining evaluation is similar. In the end, we find two diagrams contribute equally to give
\begin{align}\label{Pi_ra_final}
i\P_{\m\n}^{ra}(Q)=2e^2\int\frac{d^3p}{(2\p)^3}\frac{1}{E_p}(f_e(\vp)+f_{\bar{e}}(\vp))\big[\frac{P_\m Q_\n+P_\n Q_\m-g_{\m\n}P\cdot Q}{P\cdot Q}-\frac{P^\m P^\n Q^2}{(P\cdot Q)^2}\big],
\end{align}
with $E_p=\sqrt{p^2+m^2}$. The Ward identity can be verified as
\begin{align}\label{ward}
iQ^\m\P_{\m\n}^{ra}(Q)=0.
\end{align}

\begin{figure}
\includegraphics[width=0.3\textwidth]{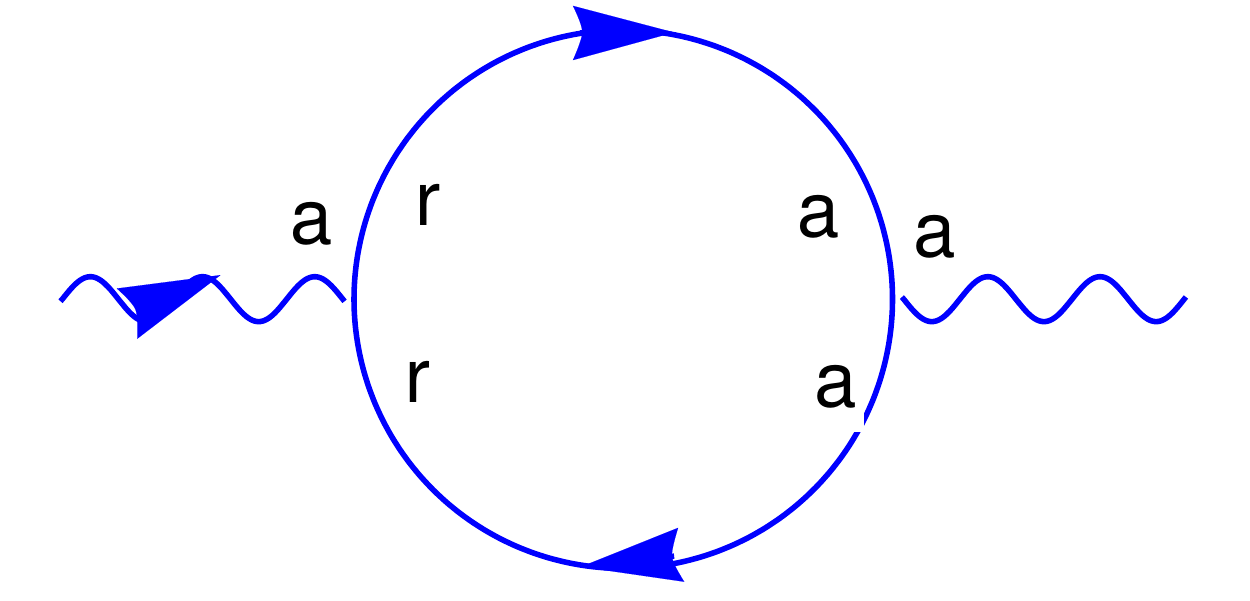}
\includegraphics[width=0.3\textwidth]{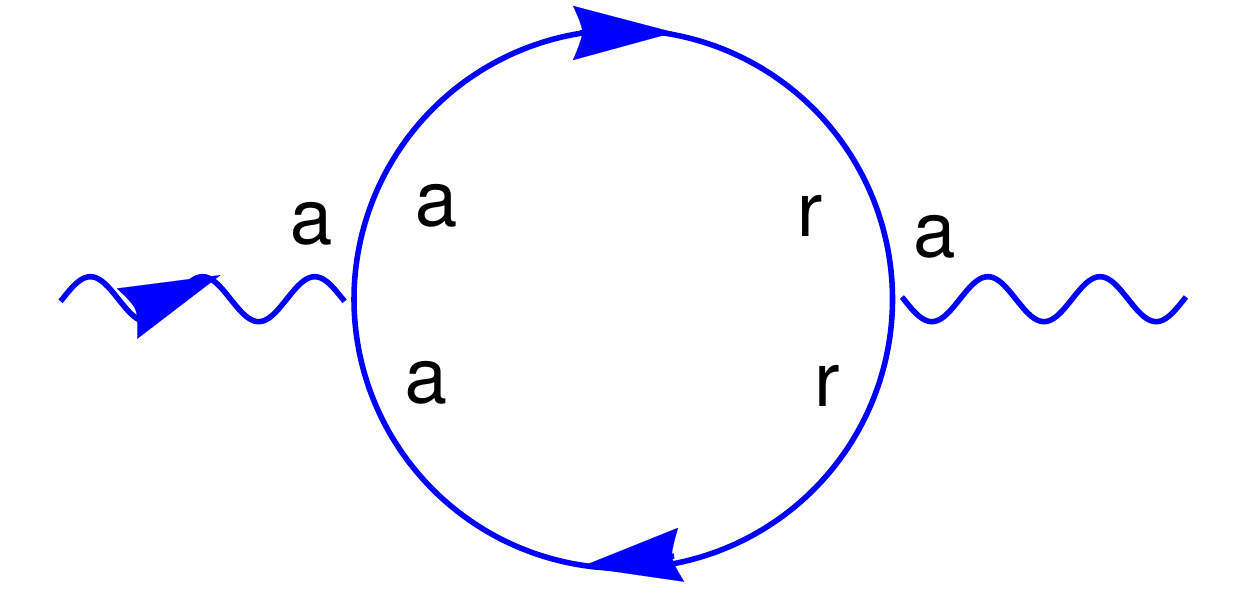}
\includegraphics[width=0.3\textwidth]{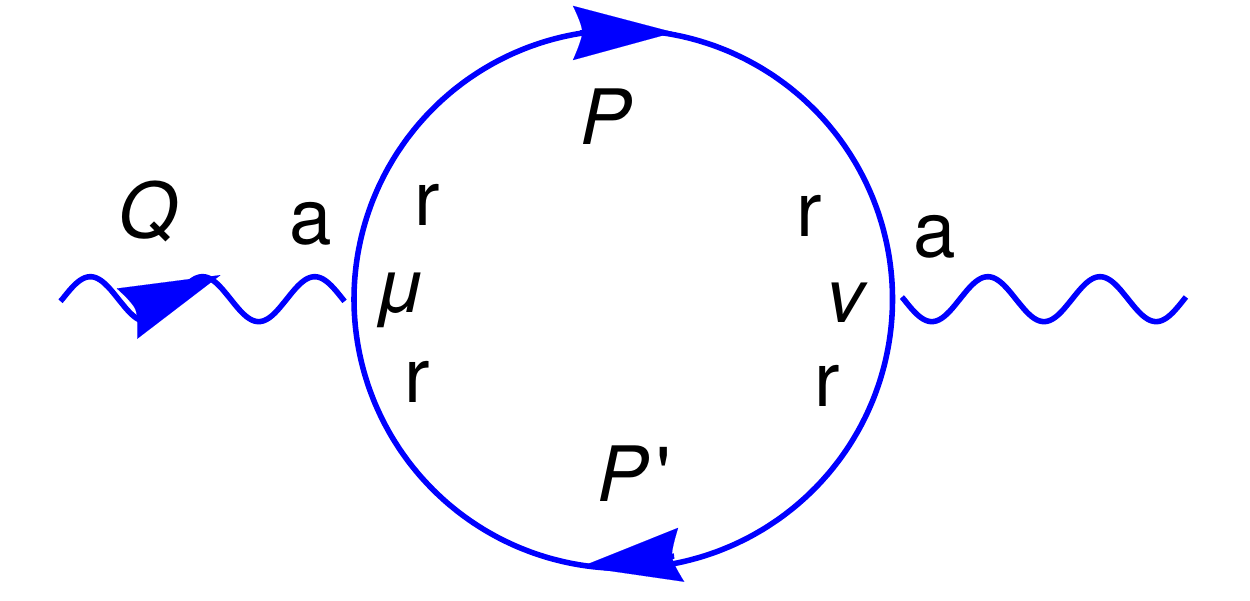}
\caption{\label{rr-photon}One-loop soft photon self-energy $\P_{aa}$. Vertices can have labelings $rra$ and $aaa$. Only the third diagram is medium dependent.}
\end{figure}

Finally we calculate the self-energy $\P_{\m\n}^{aa}$ for soft photon, which will be used to determine $D_{\m\n}^{rr}$. In this case, we also have $m\lesssim e\L$ so that inelastic collisions is relevant. $\P_{\m\n}^{aa}$ contains contributions from three diagrams in Fig.~\ref{rr-photon}. Only the last one is medium dependent, which gives
\begin{align}\label{Pi_aa}
\P_{\m\n}^{aa}(Q)&=e^2\int_{P,P'}(2\p)^4\d(P-P'-Q)\tr[\g_\m\slashed{P}\g_\n\slashed{P'}]2\p\e(p_0)\d(P^2)\(\frac{1}{2}-f_e(P)\)\no
&\times 2\p\e(p_0')\d(P'{}^2)\(\frac{1}{2}-f_e(P')\).
\end{align}
Note that we have dropped mass in the propagators. 
For $Q\ll P,P'$, $p_0$ and $p_0'$ have the same sign, thus $\e(p_0)\e(p_0')=1$. We expand the delta functions and evaluate the trace to obtain
\begin{align}\label{Pi_aa_expand}
&\P_{\m\n}^{aa}(Q)=e^2\int\frac{d^3p}{(2\p)^3}\frac{1}{4pp'}(2\p)\d(p-|{\bf p}-{\bf q}|-q_0)\big[\(4P_\m P'{}_\n+4P_\n P'{}_\m-4g_{\m\n}P\cdot P'\)\no
&(\frac{1}{2}-f_e(\vp))(\frac{1}{2}-f_e(\vp\,'))
+\(4\bar{P}_\m \bar{P}'{}_\n+4\bar{P}_\n \bar{P}'{}_\m-4g_{\m\n}\bar{P}\cdot \bar{P}'\)(\frac{1}{2}-f_{\bar{e}}(-\vp))(\frac{1}{2}-f_{\bar{e}}(-\vp\,'))\big].
\end{align}
We further approximate $P'{}_\m\simeq P_\m$, $\bar{P}'{}_\m\simeq \bar{P}_\m$ and neglect $P\cdot P'=-Q^2+2m^2$. Finally we make a change of variable $\vp\to -\vp$. It amounts to the replacement $\bar{P}\to -P$ and $f_{\bar{e}}(-\vp)\to f_{\bar{e}}(\vp)$, after which we obtain
\begin{align}\label{Pi_aa_final}
\P^{aa}_{\m\n}=e^2\int\frac{d^3p}{(2\p)^3}2\hat{P}_\m\hat{P}_\n(2\p)\d(p-|{\bf p}-{\bf q}|-q_0)\(f_e(\vp)^2-f_e(\vp)\)+(f_e\to f_{\bar{e}}),
\end{align}
with $\hat{P}^\m=(1,\hat{p})$.

To determine $D^{rr}_{\m\n}$, we use the following identities
\begin{align}\label{ra_12}
&D^{rr}_{\m\n}=\frac{1}{2}\(D_{\m\n}^>+D_{\m\n}^<\),\no
&\P^{aa}_{\m\n}=\frac{1}{2}\(\P_{\m\n}^>+\P_{\m\n}^<\).
\end{align}
Using $D_{\m\n}^{</>}=-D^R_{\m\a}\P^{\a\b</>}D^A_{\b\n}+O(\hb)$, we obtain
\begin{align}\label{Drr}
D_{\m\n}^{rr}=-D^R_{\m\a}\P_{aa}^{\a\b}D^A_{\b\n},
\end{align}
where $D^R$ is given by \eqref{photon_R} and $D^A$ determined from \eqref{hermiticity}.

\section{Damping rate of hard fermion}

The damping rate $\G$ is determined from the dispersion $p_0=E_p-\frac{i\G}{2}$, which is the root of the following equation
\begin{align}\label{disp2}
(P_\m-\S^R_\m)^2-(m-\S^R_m)^2=0,
\end{align}
with $\S^R=\S^R_\m\g^\m+\S^R_m1$.
It is the same equation as \eqref{disp}, but now we need to find out the imaginary part of $p_0$. Note that $\S_\m^R\sim e^2$ and $\S_m^R=0$ to the order of our interest. From \eqref{disp2}, we obtain
\begin{align}\label{Gamma}
\G=-\frac{1}{2p_0}\text{Im}\(\tr[\slashed{P}\S^R]\)=-\frac{1}{2p_0}\text{Re}\(\tr[\slashed{P}\S^{ra}]\).
\end{align}
We proceed with the following representation from the left panel of Fig.~\ref{Rse-fermion} (with $P$ and $Q$ exchanged)
\begin{align}\label{Sra_rep}
\S^{ra}(P)=-e^2\int_Q\g^\m S_{ar}(P-Q)\g^\n D_{\n\m}^{rr}(Q).
\end{align}
Plugging \eqref{Sra_rep} into \eqref{Gamma} and using that $D_{\m\n}^{rr}$ is real, we obtain
\begin{align}
\text{Re}(\tr[\slashed{P}\S^{ra}])=-e^2\int_Q\text{Re}\(\tr[\slashed{P}\g^\m(\slashed{P}-\slashed{Q})\g^\n]\frac{i}{(p_0-q_0-i\e)^2-(|{\bf p}-{\bf q}|)^2-m^2}\)D_{\n\m}^{rr}(Q).
\end{align}
Since $Q\ll P$, we can approximate $\tr[\slashed{P}\g^\m(\slashed{P}-\slashed{Q})\g^\n]\simeq 8P^\m P^\n$ and take the real part as
\begin{align}
&\text{Re}\(\frac{i}{(p_0-q_0-i\e)^2-(|{\bf p}-{\bf q}|)^2-m^2}\)\simeq\text{Re}\(\frac{i}{2p_0q_0-2\vp\cdot\vq-p_0i\e}\)\no
=&\d(2p_0q_0-2\vp\cdot\vq)\e(p_0)=\frac{\p}{2p_0}\d(q_0-q_\pr),
\end{align}
where $q_\pr=\vq\cdot\hat{p}$.
In the end, we have
\begin{align}
\G=e^2\int\frac{d^3q}{(2\p)^3}\hat{P}^\m \hat{P}^\n D_{\n\m}^{rr}(Q).
\end{align}
Since $Q$ is soft, the suppression factor $e^3$ from phase space $d^3q$ is compensated by the Bose enhanced propagator $D^{rr}\sim e^{-3}$, giving $\G\sim e^2$. Such enhancement mechanism is not present in the contribution from right panel of Fig.~\ref{Rse-fermion}, so we ignore.

\section{Relating four-point correlators in $12$ and $ra$ basis}

The lesser and greater photon self-energies require knowledge of fermion four-point correlators in the $12$ basis, while we calculate in Section \ref{sec_inec} only one particular correlator in the $ra$ basis. In this appendix, we establish a relation between them. The fermion four-point correlator is defined as
\begin{align}\label{4pt}
G_{ijkl}(x,x',y,y')=\lag\ps_i(x)\bar{\ps}_k(y)\ps_l(y')\bar{\ps}_j(x')\rag,
\end{align}
with labels taking values either in $12$ or in $ra$ basis.
%
Using the basic relation between field in different basis
\begin{align}
\ps_r=\frac{1}{2}(\ps_1+\ps_2),\quad \ps_a=\ps_1-\ps_2,
\end{align}
we easily obtain
\begin{align}\label{4pt_1122}
G_{1122}=G_{rrrr}+\frac{1}{2}G_{rarr}+\frac{1}{2}G_{arrr}-\frac{1}{2}G_{rrar}-\frac{1}{2}G_{rrra}+\frac{1}{4}G_{aarr}+\frac{1}{4}G_{rraa}+\cdots,
\end{align}
with $\cdots$ including correlators with at least three $a$'s in $ra$ basis. They are not allowed in the pinching kinematical region\footnote{This is because in these cases at least one end is labeled by two $a$'s, then the opposite end is forced to have two $r$'s, contradicting the assumed number of $a$'s.}. Note that we have calculated $G_{aarr}$. The other labelings appearing in \eqref{4pt_1122} can be simply related to $G_{aarr}$ as we now set out to find.

\begin{figure}
\includegraphics[width=0.8\textwidth]{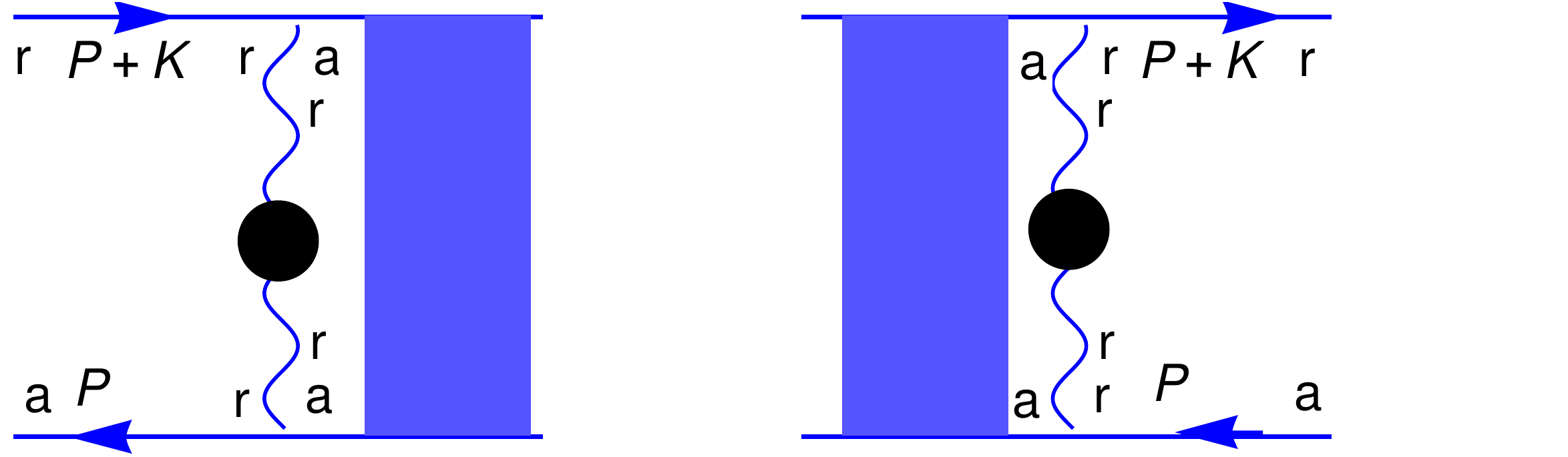}
\caption{\label{3r1a}Labelings for $G_{rarr}$ and $G_{rrra}$. By pinching mechanism, the labels are uniquely fixed. $G_{arrr}$ and $G_{rrar}$ can be obtained by exchanging $r$ and $a$ on one end. The shaded structures corresponds to $G_{aarr}$(left panel) and $G_{rraa}$(right panel) respectively.}
\end{figure}
From the diagrammatic representations in Fig.\ref{3r1a}, we easily deduce using $S^{rr}=\(S^{ra}-S^{ar}\)\(\frac{1}{2}-f_e\)$
\begin{align}\label{some_kms}
&G_{rarr}=-\(\frac{1}{2}-f_e(P+K)\)G_{aarr},\no
&G_{rrra}=\(\frac{1}{2}-f_e(P+K)\)G_{rraa},\no
&G_{arrr}=\(\frac{1}{2}-f_e(P)\)G_{aarr},\no
&G_{rrar}=-\(\frac{1}{2}-f_e(P)\)G_{rraa},
\end{align}
where we have extracted relevant component of $S^{rr}$ in arriving at the above. For example, we keep $S^{ar}$ in $G_{rarr}$ dropping $S^{ra}$ because it is paired with another $S_{ra}$.
\begin{figure}
\includegraphics[width=0.8\textwidth]{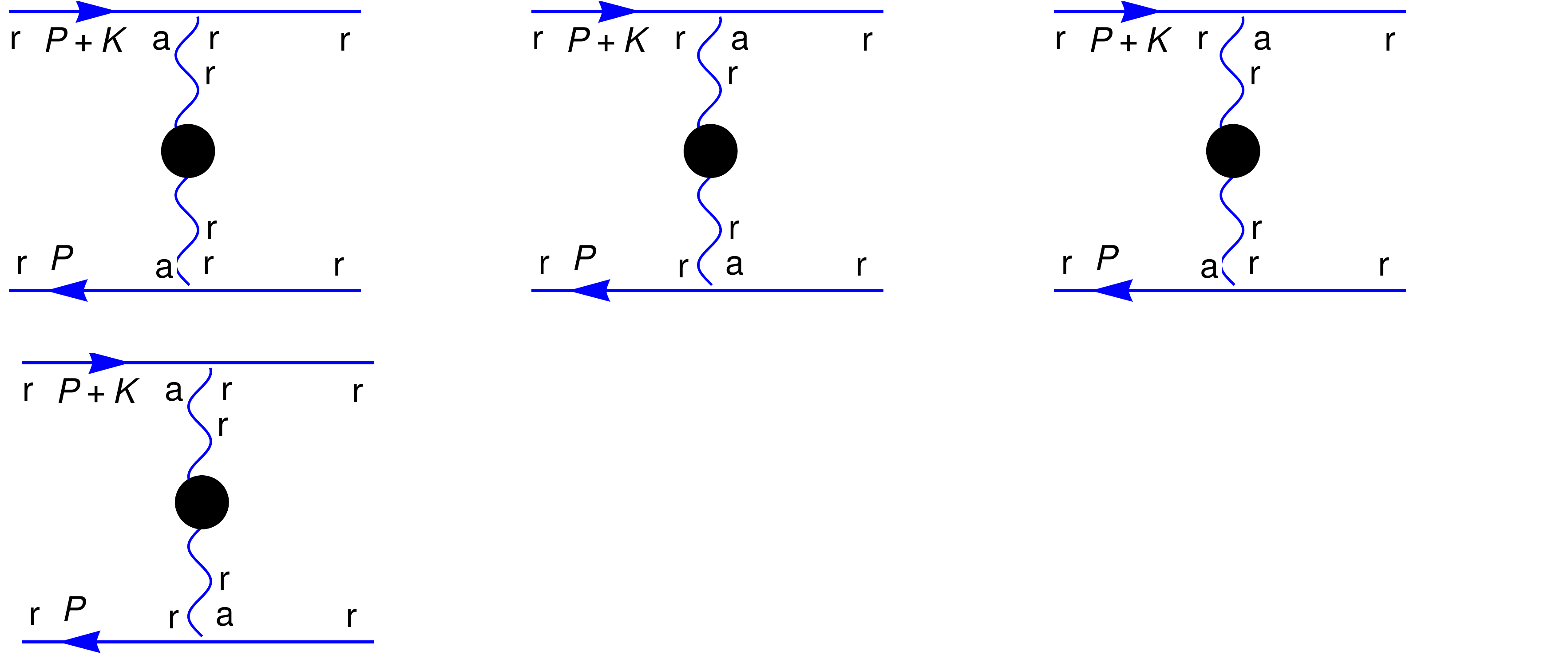}
\caption{\label{alternate}Possible subdiagrams contributing to alternating unit $\(\begin{smallmatrix}S_{ra}(P+K)&S_{ar}(P+K)\\S_{ar}(P)&S_{ra}(P)\end{smallmatrix}\)$. Note that the labels for the part extending from both sides of the subdiagrams are uniquely fixed.}
\end{figure}
$G_{rrrr}$ is a little complicated. We claim contributions containing alternating propagators like $\(\begin{smallmatrix}S_{ra}(P+K)&S_{ar}(P+K)\\S_{ar}(P)&S_{ra}(P)\end{smallmatrix}\)$ all vanish. We show below such contribution cancel among different diagrams.
Referring to Fig.~\ref{alternate}, we see the alternating contribution can arise from four possible subdiagrams. Extracting the proportionality function of $\(\begin{smallmatrix}S_{ra}(P+K)&S_{ar}(P+K)\\S_{ar}(P)&S_{ra}(P)\end{smallmatrix}\)$, we have 
\begin{align}
2\(\frac{1}{2}-f_e(P)\)\(\frac{1}{2}-f_e(P+K)\)-2\(\frac{1}{2}-f_e(P)\)\(\frac{1}{2}-f_e(P+K)\)=0.
\end{align}

Therefore only non-alternating contributions are allowed, which are either $G_{aarr}$ or $G_{rraa}$. It follows
\begin{align}\label{half_kms}
G_{rrrr}&=-\(\frac{1}{2}-f_e(P)\)\(\frac{1}{2}-f_e(P+K)\)(G_{aarr}+G_{rraa})\no
&=-\(\frac{1}{2}-f_e(P)\)\(\frac{1}{2}-f_e(P+K)\)2\text{Re}(G_{rraa}),
\end{align}
where we have used $G_{rraa}=G_{aarr}^*$.
%
Similar analysis gives
\begin{align}
G_{2211}=f_e(P)(1-f_e(P+K))2\text{Re}G_{aarr}.
\end{align}
They are off-equilibrium generalization of KMS relations in \cite{Wang:1998wg} in the special pinching kinematical region.

\begin{figure}
\includegraphics[width=0.8\textwidth]{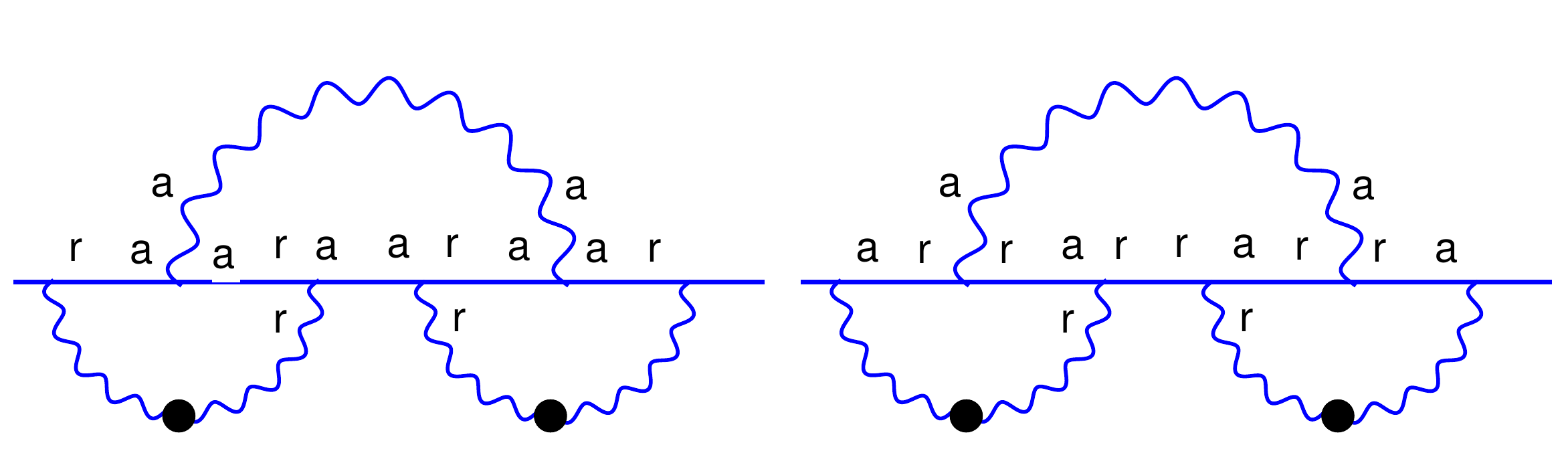}
\includegraphics[width=0.8\textwidth]{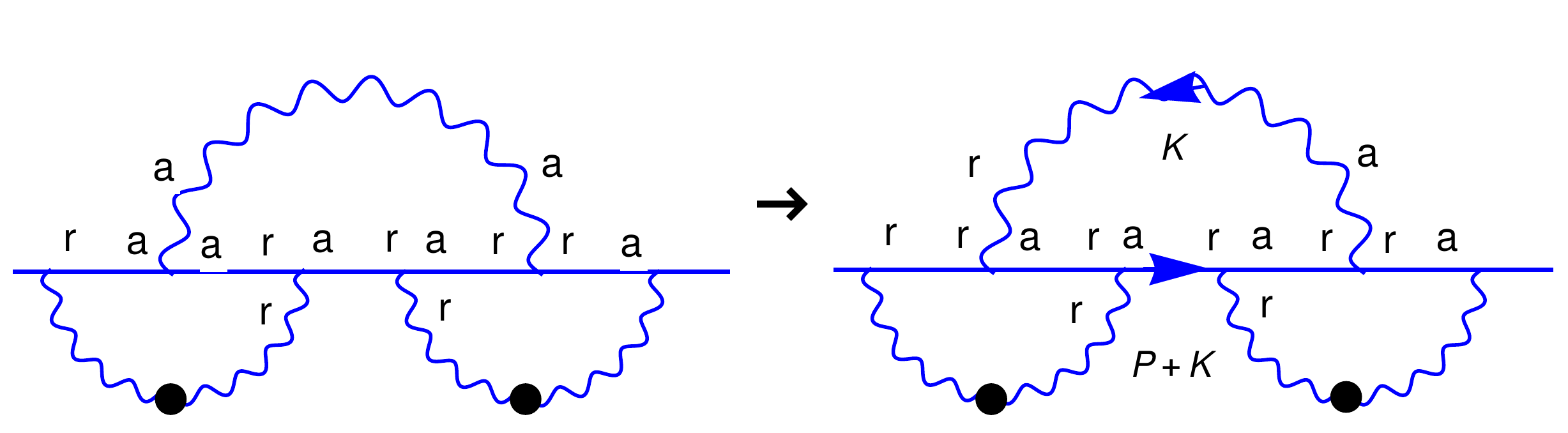}
\caption{\label{2vertices}The top row shows self-energy diagrams with both vertices constructed from $G_{rraa}$ or $G_{aarr}$. These diagrams contain one $aa$ fermion propagator and one $aa$ photon propagator. By flipping labels in $G_{rraa}$ or $G_{aarr}$, There is always one remaining $aa$ propagator. The bottom row shows a self-energy diagram with vertices constructed from $G_{rraa}$ and $G_{aarr}$ on the left. A possible flipping without $aa$ propagator is shown on the right, which vanishes upon integration of $k_0$. Multiple soft photon exchanges are allowed on each vertex. Only the outermost one is shown for clarity.}
\end{figure}

\section{Fermion self-energies with both vertices resummed}

In this appendix, we show fermion self-energy with both vertices resummed is not allowed. We work in the $ra$ basis. Recall from the previous appendix that all the non-vanishing four point correlators can be generated from the two basic ones $G_{aarr}$ and $G_{rraa}$ by flipping labels from $a$ to $r$. We show in Fig.~\ref{2vertices} three inequivalent diagrams with two vertices constructed using $G_{aarr}$ and $G_{rraa}$. Other diagrams can be generated from them by flipping labels. Since labels can only be flipped from $a$ to $r$, we find diagrams in the top row and those with labels flipped always contain at least an $aa$ propagator for either fermion or photon, thus are not allowed. In the bottom row, we do find a possible flipping without $aa$ propagators. However, this diagram contains the pair of propagators $S_{ar}(P+K)D^{\m\n}_{ar}(K)\sim D_{ar}(P+K)D_{ar}(K)$, which vanishes identically upon integration of $k_0$ by closing the contour properly.

\bibliographystyle{unsrt}
\bibliography{QED_AKT}

\end{document}